\documentclass[journal,twocolumn,10pt,twoside]{IEEEtran}
\normalsize

\ifCLASSINFOpdf
   \usepackage[pdftex]{graphicx}
   \usepackage{epstopdf}
\else
\fi

\usepackage{url}
\usepackage{amsmath}
\usepackage{amssymb}
\usepackage{cite}
\usepackage{graphicx}
\usepackage{algorithm}
\usepackage{algpseudocode}
\usepackage{subfigure}
\usepackage{multirow} 
\usepackage{longtable}
\usepackage{threeparttable}
\usepackage{caption3}
\usepackage{array}
\usepackage{cases}
\usepackage{color}
\usepackage[T1]{fontenc}
\usepackage[utf8]{inputenc}
\usepackage{authblk}
\usepackage{subfigmat}

\newcommand{\rf}{\mathrm{rf}}
\newcommand{\dc}{\mathrm{dc}}
\newcommand{\In}{\mathrm{in}}
\newcommand{\out}{\mathrm{out}}
\newcommand{\ant}{\mathrm{ant}}

\newtheorem{remark}{Remark}

\begin{document}
\title{On the Beneficial Roles of Fading and Transmit Diversity in Wireless Power Transfer with Nonlinear Energy Harvesting}
\author{Bruno Clerckx and Junghoon Kim 
\thanks{B. Clerckx and J. Kim are with the EEE department at Imperial College London, London SW7 2AZ, UK (email: \{b.clerckx,junghoon.kim15\}@imperial.ac.uk). This work has been partially supported by the EPSRC of the UK under grant EP/P003885/1.
}
}

\maketitle

\begin{abstract} We study the effect of channel fading in Wireless Power Transfer (WPT) and show that fading enhances the RF-to-DC conversion efficiency of nonlinear RF energy harvesters. We then develop a new form of signal design for WPT, denoted as Transmit Diversity, that relies on multiple dumb antennas at the transmitter to induce fast fluctuations of the wireless channel. Those fluctuations boost the RF-to-DC conversion efficiency thanks to the energy harvester nonlinearity. In contrast with (energy) beamforming, Transmit Diversity does not rely on Channel State Information at the Transmitter (CSIT) and does not increase the average power at the energy harvester input, though it still enhances the overall end-to-end power transfer efficiency. Transmit Diversity is also combined with recently developed (energy) waveform and modulation to provide further enhancements. The efficacy of the scheme is analyzed using physics-based and curve fitting-based nonlinear models of the energy harvester and demonstrated using circuit simulations, prototyping and experimentation. Measurements with two transmit antennas reveal gains of 50\% in harvested DC power over a single transmit antenna setup. The work (again) highlights the crucial role played by the harvester nonlinearity and demonstrates that multiple transmit antennas can be beneficial to WPT even in the absence of CSIT.
\end{abstract}
\vspace{-0.2cm}

\IEEEpeerreviewmaketitle

\section{Introduction}
\IEEEPARstart{F}{ading} and diversity have been studied for several decades in wireless communications and are at the heart of any modern wireless communication system design. Channel fading is known to be a major challenge that affects the performance of wireless communication systems. Diversity techniques have been designed to enhance the reliability of the transmission over fading channels by providing and combining multiple independent replicas (in space, frequency or time) of the transmit signal \cite{Tse:2005}. In modern communication systems, spatial diversity is commonly exploited through the use of multiple antennas at the transmitter and the receiver. Of particular interest, is transmit diversity that exploits spatial diversity at a multi-antenna transmitter without the need for any Channel State Information at the Transmitter (CSIT). This contrasts with transmit beamforming that relies on CSIT to focus energy and provide an array/beamforming gain.   
\par Wireless Power Transfer (WPT) via radio-frequency radiation is nowadays regarded as a feasible technology for energising low-power devices in Internet-of-Things (IoT) applications. The major challenge with WPT, and therefore any wireless power-based system such as Simultaneous Wireless Information and Power Transfer (SWIPT), Wirelessly Powered Communication Network (WPCN) and Wirelessly Powered Backscatter Communication (WPBC), is to find ways to increase the end-to-end power transfer efficiency, or equivalently the DC power level at the output of the rectenna for a given transmit power. To that end, the traditional line of research in the RF literature has been devoted to the design of efficient rectennas \cite{OptBehaviour,Costanzo:2016}, but a new line of research on communications and signals design for WPT has emerged recently in the communication literature \cite{Zeng:2017}. Various signal strategies have\:been\:developed\:to\:boost\:the\:harvested DC power. 
\par A \textit{first} strategy is to design (energy) \textit{beamforming} so as to increase the RF input power to the energy harvester or equivalently increase the RF-to-RF transmission efficiency $e_{\rf-\rf}$ \cite{Zeng:2017}. The simplest beamformer is obtained by Maximum Ratio Transmission (MRT), similarly to wireless communications. Multi-antenna beamforming relies on CSIT, and therefore requires appropriate channel acquisition strategies specifically designed for WPT \cite{Xu_Zhang:2014}. Alternative techniques to multi-antenna beamforming, also enabling directional/energy focusing transmission, consist in time-modulated arrays and retrodirective arrays \cite{Costanzo:2016} and time-reversal techniques \cite{Ku:2017}. 
\par A \textit{second} strategy is to design (energy) \textit{waveform} so as to exploit the nonlinearity of the energy harvester and increase the RF-to-DC conversion efficiency $e_{\rf-\dc}$ \cite{Trotter:2009,Clerckx:2016b}. A systematic design of waveform for WPT was first proposed in \cite{Clerckx:2016b}. Waveforms can be designed with and without CSIT depending on the frequency selectivity of the channel. In frequency-selective channels, a channel-adaptive waveform exploits jointly the channel frequency selectivity and the energy harvester nonlinearity so as to maximize $e_{\rf-\rf} \times e_{\rf-\dc}$. Waveforms can also be designed for a multi-antenna transmitter so as to additionally exploit a beamforming gain \cite{Clerckx:2016b,Huang:2017}. Channel-adaptive waveforms rely on CSIT and therefore require appropriate channel acquisition strategies, suited for WPT with nonlinear energy harvesting \cite{Huang:2018}. Due to the harvester nonlinearity, the waveform design commonly results from a non-convex optimization problem, which does not lend itself easily to implementation. Strategies to decrease the waveform design complexity have further appeared in \cite{Huang:2017,Clerckx:2017,Moghadam:2017}. Though the design commonly assumes deterministic multisine (i.e. unmodulated multi-carrier) waveforms, modulated (single-carrier and multi-carrier) waveforms have also recently appeared, especially in the context of wireless information and power transfer \cite{Clerckx:2018b}. 
\par A \textit{third} strategy is to design (energy) \textit{modulation} for single-carrier transmission so as to exploit the nonlinearity of the energy harvester and increase the RF-to-DC conversion efficiency $e_{\rf-\dc}$ \cite{Varasteh:2017,Varasteh:2017b}. As explained in \cite{Clerckx:2018b}, modulated and deterministic signals do not lead to the same RF-to-DC conversion efficiency due to the energy harvester nonlinearity. The design of modulation specifically aimed for WPT has recently emerged. It was shown in \cite{Varasteh:2017,Varasteh:2017b} that a real Gaussian input leads to a higher harvested DC power than a circularly symmetric complex Gaussian input. The best known modulation for WPT, developed in \cite{Varasteh:2017b}, exhibits a low probability of high amplitude signal, reminiscent of flash-signaling. Modulation can also be combined with multi-antenna so as to benefit from an additional beamforming gain. 
\par Note that, in contrast with the beamforming strategy, the waveform and modulation strategies are deeply rooted in the nonlinearity of the energy harvester. The energy harvester nonlinearity has now appeared to play a crucial role in WPT signal and system design but also in various wireless information and power transfer signal and system designs \cite{Clerckx:2018c}.
\par In this paper, in view of the fundamental role played by fading and diversity in wireless communications, and the key role played by nonlinearity in WPT, we ask ourselves the following questions: \textit{Is fading beneficial or detrimental to WPT?} \textit{Can diversity play a role in WPT?} \textit{Can multiple transmit antennas be useful to WPT in the absence of CSIT?} At first, if we consider a naive way of modeling the energy harvester using the so-called linear model, which assumes a constant RF-to-DC conversion efficiency independent of the harvester's input signal (power and shape), maximizing the output DC power is equivalent to maximizing the average RF input power to the energy harvester \cite{Zeng:2017,Clerckx:2016b}. Since fading does not increase the average RF input power, we may be tempted to answer that fading does not impact the harvested DC power. Similarly, since multiple transmit antennas do not increase the average RF input power in the absence of CSIT, we may be tempted to answer that multiple antennas are useless in the absence of CSIT. Following the same line of thoughts, diversity would be useless either since fading would not have any impact on the harvested DC performance. In this paper, we show that, \textit{on the contrary}, as a consequence of the energy harvester nonlinearity, fading is beneficial to the harvested DC power, multiple transmit antennas are useful even in the absence of CSIT and diversity has a role to play in WPT. This is the first paper to make those observations. Specifically, the contributions of this paper are summarized as follows.
\par \textit{First}, using two different models of the energy harvester nonlinearity, namely the physics-based diode model of \cite{Clerckx:2016b} and a curve fitting-based model approach similar to \cite{Boshkovska:2015,Xu:2017,Alevizos:2018}, we study the impact of channel fading on the harvested DC power and highlight that channel fading is beneficial to WPT. 
\par \textit{Second}, aside the three aforementioned beamforming, waveform and modulation strategies for WPT, we develop a new (and therefore a \textit{fourth}) WPT signal strategy, denoted as transmit diversity, that relies on multiple dumb antennas at the transmitter to induce fast fluctuations of the wireless channel. Those fluctuations are shown to boost the RF-to-DC conversion efficiency and the harvested DC power thanks to the energy harvester nonlinearity. Transmit diversity does not rely on any form of CSIT and its performance analysis highlights that multiple transmit antennas are useful even in the absence of CSIT. We also show that transmit diversity can be combined with the (energy) waveform and modulation to provide further enhancements. The gain of transmit diversity is analyzed using the two aforementioned nonlinear models.  
\par \textit{Third}, the analysis and performance benefits are confirmed using realistic circuit simulations, prototyping and experimentation. Real-time over-the-air measurements reveal gains of about 50\% to 100\% in harvested DC power with a two-antenna transmit diversity strategy over single-antenna setup, without any need for CSIT. Transmit diversity is well suited for massive IoT for which CSIT acquisition is unpractical. 
\par As a main takeaway observation, results (again) highlight the crucial role played by the harvester nonlinearity in WPT signal/system design and demonstrate the previously unknown fact that multiple transmit antennas can be beneficial to WPT even in the absence of CSIT. This opens the door to a re-thinking of the role played by multiple antennas in wireless power-based network design.

\par \textit{Organization:} Section \ref{nonlinear_model} introduces the physics-based diode nonlinear model, Section \ref{Fading_section} analyzes the impact of fading on harvested DC power, Section \ref{basic_TD_section} introduces the basic idea of transmit diversity for WPT and analyzes its performance, Section \ref{section_TD_mod_wf} shows that transmit diversity can be combined with (energy) waveform and modulation to provide further performance benefits, Section \ref{nonlinear_model_fit_section} revisits the impact of fading and transmit diversity using a curve fitting-based nonlinear model, Section \ref{simulations} provides circuit simulation, prototyping and experimentation results, and Section \ref{conclusions} concludes the work and discusses future research directions.

\par \emph{Notations:} In this paper, scalars are denoted by italic letters. Boldface lower- and upper-case letters denote vectors and matrices, respectively. $j$ denotes the imaginary unit, i.e., $j^2=-1$. $\mathbb{E}[\cdot]$ denotes statistical expectation and $\Re\{\cdot\}$ represents the real part of a complex number. $|.|$ and $\left\|.\right\|$ refer to the absolute value of a scalar and the 2-norm of a vector. In the context of random variables, i.i.d. stands for independent and identically distributed. The distribution of a Circularly Symmetric Complex Gaussian (CSCG) random variable with zero-mean and variance $\sigma^2$ is denoted by $\mathcal{CN}(0,\sigma^2)$; hence with the real/imaginary part distributed as $\mathcal{N}(0,\sigma^2/2)$. The distribution of an exponential random variable with parameter $\lambda$ is denoted by $\mathrm{EXPO}(\lambda)$. $\sim$ stands for ``distributed as''. $\log$ is in base $e$. $\stackrel{N\nearrow}{\approx}$ means approximately equal as $N$ grows large. 
\vspace{-0.5cm}
\section{The Nonlinear Rectenna Model}\label{nonlinear_model}
\begin{figure}
   \centerline{\includegraphics[width=3.3 in]{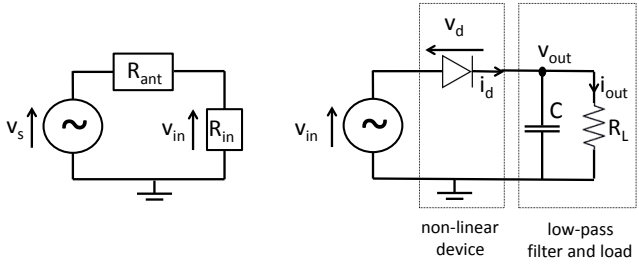}}
  \caption{Antenna equivalent circuit (left) and a single diode rectifier (right) \cite{Clerckx:2016b}. The rectifier comprises a non-linear device (diode) and a low-pass filter (consisting of a capacitor C and a load R\textsubscript{L}).}
  \label{antenna_model}
\end{figure}

We assume the same rectenna model as in \cite{Clerckx:2016b}. The rectenna is made of an antenna and a rectifier. The antenna model reflects the power transfer from the antenna to the rectifier through the matching network. A lossless antenna is modeled as a voltage source $v_{\mathrm{s}}(t)$ followed by a series resistance\footnote{Assumed real for simplicity. A more general model can be found in \cite{Clerckx:2017}.} $R_{\ant}$ (Fig. \ref{antenna_model} left hand side). Let $Z_{\In}\!=\! R_{\In} + j X_{\In}$ denote the input impedance of the rectifier and the matching network. Let $y(t)$ also denote the RF signal impinging on the receive antenna. Assuming perfect matching ($R_{\In} = R_{\ant}$, $X_{\In} = 0$), the available RF power $P_{\rf}$ is transferred to the rectifier and absorbed by $R_{\In}$, so that $P_{\rf} = \mathbb{E}\big[y(t)^2\big]\!=\!\mathbb{E}\big[v_{\In}(t)^2\big]/R_{\In}$, $v_{\In}(t)\!=\!v_{\mathrm{s}}(t)/2$, and $v_{\In}(t)\!=\!y(t)\sqrt{ R_{\In}}\!=\!y(t)\sqrt{ R_{\ant}}$. We also assume that the antenna noise is too small to be harvested.

\par We consider for simplicity a rectifier composed of a single diode followed by a low-pass filter with a load ($R_{\mathrm{L}}$), as illustrated in Fig. \ref{antenna_model} (right hand side). We consider this setup as it is the simplest rectifier configuration. Nevertheless the model presented in this subsection is not limited to a single series diode but also holds for more general rectifiers with many diodes as shown in \cite{Clerckx:2017}. Denote the voltage drop across the diode as $v_{\mathrm{d}}(t)=v_{\In}(t)-v_{\out}(t)$ where $v_{\In}(t)$ is the input voltage to the diode and $v_{\out}(t)$ is the output voltage across the load resistor. A tractable behavioral diode model is obtained by Taylor series expansion of the diode characteristic function
\begin{equation}
i_{\mathrm{d}}(t)=i_{\mathrm{s}} \big(e^{\frac{v_{\mathrm{d}}(t)}{n v_{{\mathrm{t}}}}}-1 \big),
\end{equation}
with the reverse bias saturation current $i_{{\mathrm{s}}}$, the thermal voltage $v_{{\mathrm{t}}}$, the ideality factor $n$ assumed to be equal to $1.05$, around a quiescent operating point $v_{\mathrm{d}}(t)=a$. We have
\begin{equation}\label{DiodeCurrent}
i_{\mathrm{d}}(t)=\sum_{i=0}^{\infty }k_i' \left(v_{\mathrm{d}}(t)-a\right)^i,
\end{equation}
where $k_0'=i_{\mathrm{s}}\big(e^{\frac{a}{n v_{\mathrm{t}}}}-1\big)$ and $k_i'=i_{\mathrm{s}}\frac{e^{\frac{a}{n v_{\mathrm{t}}}}}{i!\left(n v_{\mathrm{t}}\right)^i}$, $i=1,\ldots,\infty$.

\par Choosing\footnote{We here assume a steady-state response and an ideal rectification. The low pass filter is ideal such that $v_{\out}(t)$ is at constant DC level $v_{\out}$ (we drop the dependency on $t$) and the output current is also at constant DC level $i_{\out}$.} $a=\mathbb{E}[ v_{\mathrm{d}}(t) ]=-v_{\out}$, we can write  
\begin{equation}\label{current_expansion}
i_{\mathrm{d}}(t)=\sum_{i=0}^{\infty }k_i' v_{\In}(t)^i=\sum_{i=0}^{\infty }k_i' R_{\ant}^{i/2} y(t)^i.
\end{equation} 
The DC current delivered to the load and the harvested DC power are then given by
\begin{equation}\label{diode_current_power}
i_{\out}=\mathbb{E}[i_{\mathrm{d}}(t)], \hspace{1cm} P_{\dc}=i_{\out}^2 R_{\mathrm{L}},
\end{equation}
respectively. Note that the operator $\mathbb{E}[\cdot]$ takes the DC component of the diode current $i_{\mathrm{d}}(t)$ but also averages over the potential randomness carried by the input signal $y(t)$ \cite{Clerckx:2018b}.

\par The classical linear model is obtained by truncating to the second order term. This is however inaccurate and inefficient as studied in numerous works \cite{OptBehaviour,Clerckx:2016b,Clerckx:2017,Clerckx:2018b,Boshkovska:2015}. The nonlinearity of the rectifier is characterized by the fourth and higher order terms \cite{Clerckx:2016b}. Truncating the expansion to order 4 (and therefore accounting for the nonlinearity), we can write $i_{\out}\approx k_0'+k_2' R_{\ant}\mathbb{E}\left[y(t)^2\right]+k_4' R_{\ant}^2\mathbb{E}\left[y(t)^4\right]$. Following \cite{Clerckx:2016b}, $i_{\out}$, and therefore $P_{\dc}$, are directly related to the quantity 
\begin{equation}\label{diode_model_2}
z_{\dc}=k_2 R_{\ant}\mathbb{E}\left[y(t)^2\right]+k_4 R_{\ant}^2\mathbb{E}\left[y(t)^4\right]
\end{equation}
where $k_i\!=\!\frac{i_\mathrm{s}}{i!\left(n v_\mathrm{t}\right)^i}$, $i\!=\!2,4$\footnote{Assuming $i_{\mathrm{s}}\!=\!5 \mu A$, a diode ideality factor $n\!=\!1.05$ and $v_{\mathrm{t}}\!=\!25.86 mV$, typical values are given by $k_2\!=\!0.0034$ and $k_4\!=\!0.3829$ \cite{Clerckx:2016b}.}. Hence, any increase in $z_{\dc}$ would lead to an increase in $i_{\out}$ and $P_{\dc}$. Metric \eqref{diode_model_2} (and the truncation to the fourth order term) has been used and validated by circuit simulations in \cite{Clerckx:2016b,Clerckx:2017,Clerckx:2018b}\footnote{As discussed in Remark 5 of \cite{Clerckx:2016b}, more investigations are needed to assess the usefulness of higher order terms (beyond 4).}.

\section{Effect of Fading on Harvested Energy}\label{Fading_section}
In this section, we study the effect of fading on $z_{\dc}$. To that end, we consider the simplest scenario of a transmitter with a single antenna whose transmit signal at time $t$ is given by an unmodulated continuous-wave
\begin{equation}\label{CW_model_fading}
x(t)=\sqrt{2P}\cos(w_0 t)=\sqrt{2P}\Re\big\{e^{j w_0 t}\big\}, 
\end{equation}
with $w_0$ the carrier frequency. The average transmit power is fixed to $P$, namely $\mathbb{E}\left[x(t)^2\right]=P$. Denoting the wireless channel between the transmit and receive antenna as $\Lambda^{-1/2} h$ with $\Lambda$ the path loss and $h$ the complex fading coefficient (it varies very slowly over time, hence we omit the time dependency), the received signal can be written as
\begin{equation}\label{fading_rec_sig}
y(t)=\sqrt{2P}\Re\left\{\Lambda^{-1/2} h e^{j w_0 t}\right\}.
\end{equation}
The RF power level at the energy harvester input is given by 
\begin{equation}
P_{\rf}=\mathbb{E}\big[y(t)^2\big]=\Lambda^{-1} |h|^2 P. 
\end{equation}
The average RF received power (average RF input power to the energy harvester), where the average is taken over the distribution of $h$, is given by 
\begin{equation}
\bar{P}_{\rf}=\mathbb{E}_h\left[P_{\rf}\right]=\Lambda^{-1} P, 
\end{equation}
where we assume that the fading coefficient $h$ is normalized such that $\mathbb{E}\left[|h|^2\right]=1$. Hence, fading does not change the RF input power to the energy harvester on average.

\par Let us plug \eqref{fading_rec_sig} in the rectenna model \eqref{diode_model_2}. We first consider the deterministic case where $h$ does not change in time and is given by $|h|^2=1$, such that $P_{\rf}=\Lambda^{-1} P=\bar{P}_{\rf}$. Recalling the analysis in \cite{Clerckx:2016b}, we can write  
\begin{equation}
z_{\dc,\mathrm{cw}}=k_2 R_{\ant}\bar{P}_{\rf}+\frac{3}{2} k_4 R_{\ant}^2 \bar{P}_{\rf}^2.\label{z_dc_no_fading}
\end{equation}
We now consider the same unmodulated continuous-wave transmission but over a fading channel, with $h$ modeled as CSCG with $\mathbb{E}\left[|h|^2\right]=1$. According to \cite{Clerckx:2016b}, after taking the expectation over the fading process, we have 
\begin{align}
\bar{z}_{\dc,\mathrm{cw}}&=\mathbb{E}_{h}\left[z_{\dc,\mathrm{cw}}\right],\nonumber\\
&=k_2 R_{\ant}\bar{P}_{\rf}\mathbb{E}\left[|h|^2\right]+\frac{3}{2} k_4 R_{\ant}^2 \bar{P}_{\rf}^2\mathbb{E}\left[|h|^4\right],\nonumber\\
&=k_2 R_{\ant}\bar{P}_{\rf}+3 k_4 R_{\ant}^2 \bar{P}_{\rf}^2\label{z_dc_fading}
\end{align}
because $\mathbb{E}\left[|h|^4\right]=2$. We note that the second order term has not changed between \eqref{z_dc_no_fading} and \eqref{z_dc_fading}, while the fourth order term in the fading case is twice as large as in the deterministic case\footnote{Higher order terms in the Taylor expansion would lead to terms proportional to $\mathbb{E}\left[|h|^i\right]$, $i=6,8,...$. Since the moments of Gaussian distribution are unbounded as the exponent goes to infinity, the harvested energy would scale to infinity if we account for all terms in the Taylor expansion. However, recall that the assumption behind the rectenna model is that the diode operates in the nonlinear region, as detailed in Section III-B and Remark 4 of \cite{Clerckx:2016b}.}. In other words, fading is \textit{beneficial} in WPT if the rectifier nonlinearity is modeled. This benefit originates from the rectifier nonlinearity that leads to the convexity of the expression of the diode current \eqref{current_expansion} with respect to $y(t)$. According to Jensen's inequality, this convexity leads to $\mathbb{E}\left[y(t)^4\right]\geq (\mathbb{E}\left[y(t)^2\right])^2 = \bar{P}_{\rf}^2 $ in \eqref{diode_model_2}, and $\mathbb{E}\left[|h|^4\right]\geq (\mathbb{E}\left[|h|^2\right])^2$ in \eqref{z_dc_fading}, and implies that fluctuations of the signal due to fading at the input of the rectifier boost the harvested energy. This contrasts with communication where fading is known to be \textit{detrimental} \cite{Tse:2005}. Indeed, the ergodic capacity of a fast fading channel is lower than the AWGN capacity due to the concavity of the $\log$ in the rate expression.  
\par Expression \eqref{z_dc_fading} was obtained under the assumption of a continuous wave (CW) as per \eqref{CW_model_fading}. Nevertheless, if the CW is modulated using energy modulation or replaced by an energy waveform, the factor of 2 in the fourth order term originating from fading would still appear on top of an additional gain originating from the use of energy modulation/waveform. This will be discussed in more details in Section \ref{section_TD_mod_wf}.
\vspace{-0.2cm}
\section{Basic Transmit Diversity}\label{basic_TD_section}
Motivated by the above observation that fading is helpful to the harvested energy, we design a transmit diversity scheme using dumb antennas to boost the harvested energy of rectennas that are exposed to slow fading channels. We aim at inducing channel fluctuations and creating an effective fast fading channel by using multiple transmit antennas. During a given harvesting period, the rectenna will therefore see a fast varying channel that will lead to a higher harvested energy. 
\vspace{-0.2cm}
\subsection{Phase Sweeping Transmit Diversity}\label{basic_TD_section_sweeping}
\begin{figure}
   \centerline{\includegraphics[width=3.1 in]{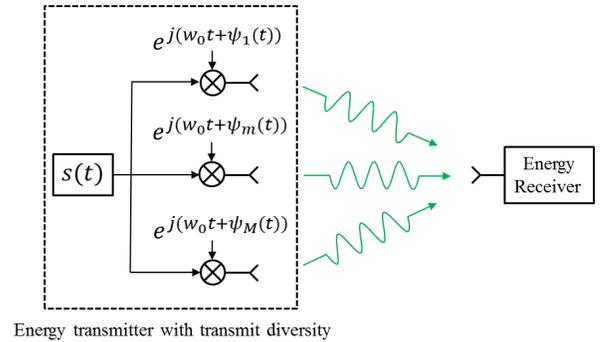}}
  \caption{General architecture of transmit diversity for WPT with $M$ dumb antennas. Signal $s(t)$ is an energy modulation/waveform ($s(t)=1$ in \eqref{TD_basic_sys_mod}).}
  \label{TD_drawing}
	\vspace{-0.3cm}
\end{figure}

In its simplest form, the proposed transmit diversity for WPT induces fast fading by transmitting a continuous wave on each antenna with an antenna dependent time-varying phase\footnote{We here assume for simplicity that the transmit power is uniformly spread across antennas and the phases only are time-varying. We could also imagine a strategy where both the amplitude and the phase on each antenna vary over time. Note also that $\psi_1(t)$ can be set to 0 without loss of generality.} $\psi_m(t)$. Considering $M$ transmit antennas and one receive antenna, we can write the transmit signal on antenna $m$ as 
\begin{align}
x_m(t)&=\sqrt{\frac{2P}{M}}\cos(w_0 t + \psi_m(t)),\nonumber\\
&=\sqrt{\frac{2P}{M}}\Re\left\{e^{j (w_0 t +\psi_m(t))}\right\}. \label{TD_basic_sys_mod}
\end{align}
The average total transmit power is kept fixed to $P$. Denoting the wireless channel between antenna $m$ and the receive antenna as $\Lambda^{-1/2} h_m$ (it varies slowly over time, hence we omit the time dependency), the received signal is written as
\begin{equation}
y(t)=\sqrt{\frac{2P}{M}}\Re\left\{\Lambda^{-1/2} h(t)e^{j w_0 t}\right\},
\end{equation}
where
\begin{equation}\label{effective_channel}
h(t)=\sum_{m=1}^M h_m e^{j \psi_m(t)}
\end{equation}
is the effective time-varying channel, whose rate of change is determined by the rate of change of the phases $\left\{\psi_m(t)\right\}_{\forall m}$. Fig. \ref{TD_drawing} illustrates the general architecture of transmit diversity.

\par
\begin{figure}
\centering
\includegraphics[width=3.1 in]{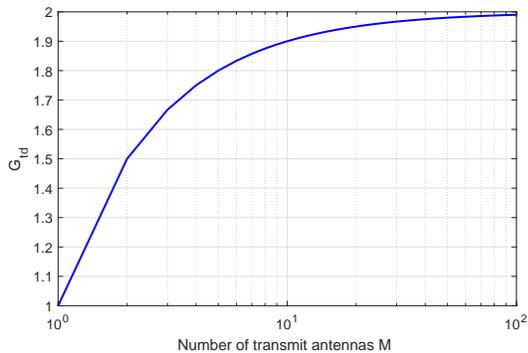}
\caption{Transmit diversity gain $G_{\mathrm{td}}$ as a function of $M$.}
\label{G_TD}
\vspace{-0.3cm}
\end{figure}

 In order to get some insights into the performance benefit of the transmit diversity strategy, we assume the phases $\left\{\psi_m(t)\right\}_{\forall m}$ are uniformly distributed over $2\pi$ and are independent. Moreover, we assume\footnote{This assumption could be experienced in strong LoS situations for instance and is made for simplicity in the analysis and circuit simulations to highlight the role of the antenna dependent time-varying phases of transmit diversity to induce fast fluctuations of a slow-varying channel.} $h_m=1$ $\forall m$. We then write
\begin{align}
\bar{z}_{\dc,\mathrm{td-cw}}&=k_2 R_{\ant}\bar{P}_{\rf}\frac{\mathbb{E}\left[|h(t)|^2\right]}{M}+\frac{3}{2} k_4 R_{\ant}^2 \bar{P}_{\rf}^2 \frac{\mathbb{E}\left[|h(t)|^4\right]}{M^2}\nonumber\\
&=k_2 R_{\ant}\bar{P}_{\rf}+\frac{3}{2} k_4 R_{\ant}^2 \bar{P}_{\rf}^2 \left(1+\frac{M-1}{M}\right).\label{zdc_td}
\end{align}
The quantity $G_{\mathrm{td}}=\left(1+\frac{M-1}{M}\right)$ is obtained from computing\footnote{The derivations are pretty straightforward and similar to the ones used to compute the scaling laws in \cite{Clerckx:2016b}. They are therefore not provided here.} $\mathbb{E}\left[|h(t)|^4\right]/M^2$ over the uniform distribution of the phases $\left\{\psi_m(t)\right\}_{\forall m}$. $G_{\mathrm{td}}$ can be viewed as the diversity gain offered by the transmit diversity strategy. It ranges from 1 for $M=1$ (i.e. no gain) to 2 in the limit of large $M$ (which corresponds to the gain offered by CSCG fading in Section \ref{Fading_section})\footnote{The diversity gain is larger when the channels on each antenna have similar strengths (e.g. $h_m=1$ $\forall m$). If there is a severe imbalance between the channel magnitudes $|h_m|$, transmit diversity will have less benefit.}, as illustrated in Fig. \ref{G_TD}. For $M=2$ and $M=4$, the scheme boosts the fourth order term by a factor of 1.5 and 1.75, respectively. This is remarkable as it shows that multiple transmit antennas are beneficial to WPT even in the absence of CSIT.

\par In a multi-user setup, transmit diversity benefits all users by providing a diversity gain to all of them simultaneously, i.e., adding more energy receivers comes for free and does not compromise each user's performance. This behavior was also observed\:with\:energy\:waveforms\:in\:frequency\:flat\:channels\:\hspace{-0.05cm}\cite{Clerckx:2016b}. 

\par It is important to recall that the transmit diversity gain is deeply rooted in the rectifier nonlinearity. If the nonlinearity modeled by the fourth order term is ignored, the transmit diversity gain disappears. With the linear model, there is no benefit in using transmit diversity since the second order term in \eqref{zdc_td} is constant irrespective of $M$ in the absence of CSIT. In other words, the second order term can only by increased by the use of beamforming, which requires CSIT. The fourth order term, on the other hand, can be enhanced by beamforming, waveform, modulation and transmit diversity, with\:the\:last\:three\:schemes\:applicable\:in\:the\:absence\:of\:CSIT\footnote{Extra performance benefits are nevertheless obtained with CSIT based-waveform design in frequency selective channels, as detailed in \cite{Clerckx:2016b}.}.

\par Transmit diversity has a number of practical benefits leading to low cost deployments. First, it relies on dumb antennas fed with a low Peak-to-Average Power Ratio (PAPR) continuous wave at a single carrier frequency $w_0$. Second, it does not require any form of synchronization among transmit antennas since a time-varying phase is applied to each of them. Third, it can be applied to co-located and distributed antenna deployments. Fourth, it is transparent to the energy receivers, which eases the system implementation. Fifth, for deployments with a massive number of devices, it provides a transmit diversity gain simultaneously to all users without any need for CSIT. 

\begin{remark}
It is worth contrasting with (energy) beamforming, waveform and modulation. Note that the transmit diversity gain $G_{\mathrm{td}}$ is obtained without any CSIT. This contrasts with a (energy) beamforming approach for WPT that always relies on CSIT in order to beam energy in the channel direction. The philosophy behind transmit diversity is also different from energy waveform \cite{Clerckx:2016b} and energy modulation \cite{Varasteh:2017b} that transmit specific signal with suitable characteristics. Energy waveform in \cite{Clerckx:2016b} relies on deterministic multisine at the transmitter whose amplitudes and phases are carefully selected (as a function of the wireless channel in general settings) so as to excite the rectifier in a periodic manner by sending high energy pulses. Energy modulation in \cite{Clerckx:2018b,Varasteh:2017,Varasteh:2017b} relies on single-carrier transmission but whose input follows a probability distribution that boosts the fourth order term in the Taylor expansion of the diode. One of such distribution, proposed in \cite{Varasteh:2017b} and reminiscent of flash signaling, exhibits a low probability of high amplitude transmit signals. In contrast, transmit diversity is designed to generate amplitude fluctuations of the wireless channel, not amplitude fluctuations of the transmit signal.  
\end{remark}

\subsection{Comparisons with Fading/Diversity in Communications}
The notion of fading and diversity has a long history in communications. It is interesting to compare the effect of fading in communications and in WPT, as well as contrast the proposed transmit diversity for WPT with various forms of diversity strategies used in communications. 
\subsubsection{Fading} Channel fading in \textit{point-to-point} (resp.\ \textit{multi-user}) scenario is commonly viewed as a source of \textit{unreliability} (resp. \textit{randomization}) in communications that has to be \textit{mitigated} (resp.\ \textit{exploited}) \cite{Viswanath:2002}. In contrast, channel fading for both \textit{point-to-point} and \textit{multi-user} scenarios can be viewed as a source of \textit{randomization} in WPT that has to be \textit{exploited}.
\subsubsection{Time Diversity} In \cite{Hiroike:1992}, a phase-rotated/sweeping transmit diversity strategy was proposed for point-to-point transmission so as to convert spatial diversity into time diversity. The effective SISO channel resulting from the addition of multiple branches seen by the receiver fades over time, and this selective fading can be exploited by channel coding. In the proposed transmit diversity for WPT, there is no information transmission and therefore no channel coding. Time-selective fading is also induced by a phase rotation/sweeping, but it is exploited by the rectifier rather than channel coding. 
\subsubsection{Multi-User Diversity} In \cite{Knopp:1995,Viswanath:2002}, multi-user diversity was introduced as a new form of diversity at the system level that leverages the independent time-varying channels across the different users. By tracking the channel fluctuations using a limited channel feedback and selecting the best user at a time (typically when its instantaneous channel is near its peak), a multi-user diversity gain is exploited that translates into a sum-rate scaling double logarithmically with the number of users at high SNR. By increasing the number of users, the sum-rate increases, though each user gets access to a smaller portion of the resources. Such a multi-user diversity strategy can also be applied in multi-user WPT in the same way as communications, as investigated in \cite{Xia:2015}. The proposed transmit diversity however differs from multi-user diversity as it does not rely on independent channels across users and does not require any form of tracking and scheduling. Moreover, all users do not have access to a portion of the resources but rather benefit simultaneously from the channel fluctuations. Therefore adding more users comes for free and does not compromise each user’s harvested DC power.
\par The proposed transmit diversity for WPT also has some similarities with multi-user diversity. Indeed, the multi-user diversity gain is known to increase with the dynamic range of the fluctuations. This may naturally occur in fast fading conditions. However, in deployments subject to Line-of-Sight (LoS) and/or slow fading, multiple transmit antennas with time-varying phases and magnitudes can be used to induce large and fast channel fluctuations, which are then exploited by channel tracking and scheduling \cite{Viswanath:2002}. Similarly, in the proposed transmit diversity for WPT, large channel fluctuations boost the harvested DC power and multiple transmit antennas with time-varying phases (and potentially magnitudes) are used to generate those channel fluctuations. 

\section{Transmit Diversity with Energy Modulation/Waveform}\label{section_TD_mod_wf}
Energy modulation and energy waveform are two different techniques to also boost the amount of DC power at the output of the rectenna. Similarly to transmit diversity, they both exploit the nonlinearity of the energy harvester. However, in contrast to transmit diversity, both create large amplitude fluctuations of the transmit signal. Energy waveform in \cite{Trotter:2009,Clerckx:2016b} relies on deterministic multisine at the transmitter so as to excite the rectifier in a periodic manner by sending high energy pulses. Energy modulation in \cite{Clerckx:2018b,Varasteh:2017,Varasteh:2017b} relies on single-carrier transmission with inputs following a probability distribution that boosts the fourth order term in the diode Taylor expansion.  
\par In this section, we show that transmit diversity can be combined with energy modulation/waveform so as to jointly generate amplitude fluctuations of the wireless channel and amplitude fluctuations of the transmit signal. This combined effect leads to a joint diversity and modulation/waveform gain and can be leveraged in practice to balance the cost of the power amplifiers (PA) versus the number of antennas. Indeed, one extreme consists in using one single transmit antenna to generate energy waveform/modulation. That would require an expensive PA to cope with the high PAPR transmit signal and would not benefit from a transmit diversity gain. Another extreme would consist in using transmit diversity with multiple dumb antennas with cheap PAs and low PAPR continuous wave signal. That would not benefit from a modulation/waveform gain. In between those two extremes, we can use of mixture of transmit diversity and energy modulation/waveform and partially trade the transmit diversity gain with the modulation/waveform gain so as to balance the cost of the PAs and the number of transmit antennas.

\subsection{Transmit Diversity with Energy Modulation}\label{subsection_TD_mod}
Rather than transmitting an unmodulated continuous wave with a time-varying phase on each antenna, we can modulate the carrier using a suitable input distribution. The transmit signal on antenna $m$ can be written as
\begin{equation}\label{TD_mod_sys_mod}
x_m(t)=\sqrt{\frac{2P}{M}}\Re\left\{s(t) e^{j (w_0 t +\psi_m(t))}\right\} 
\end{equation}
where $s(t)$ is the complex energy symbol drawn from a given distribution with $\mathbb{E}\left[|s(t)|^2\right]=1$ and transmitted on all antennas. $s(t)$ can be drawn for instance from a CSCG distribution, real distribution or any other distribution with high energy content \cite{Varasteh:2017b}. The received signal becomes
\begin{equation}\label{TD_mod_sys_mod_rec}
y(t)=\sqrt{\frac{2P}{M}}\Re\left\{\Lambda^{-1/2} h(t)s(t)e^{j w_0 t}\right\},
\end{equation}
with $h(t)$ given in \eqref{effective_channel}. In order to analyze the performance gain, let us again assume $h_m=1$. Following \cite{Clerckx:2018b}, we can then compute
\begin{align}
\bar{z}_{\mathrm{dc,td-mod}}&=k_2 R_{ant}\bar{P}_{\rf}\mathbb{E}\left\{|h(t)|^2\right\}\mathbb{E}\left\{|s(t)|^2\right\}\nonumber\\
&\hspace{1cm}+\frac{3}{2} k_4 R_{\ant}^2 \bar{P}_{\rf}^2 \mathbb{E}\left\{|h(t)|^4\right\}\mathbb{E}\left\{|s(t)|^4\right\}\nonumber\\
&=k_2 R_{\ant}\bar{P}_{\rf}+\frac{3}{2} k_4 R_{\ant}^2 \bar{P}_{\rf}^2 G_{\mathrm{td}}G_{\mathrm{mod}}.
\end{align}
where $G_{\mathrm{td}}=\left(1+\frac{M-1}{M}\right)$ and $G_{\mathrm{mod}}=\mathbb{E}\left\{|s(t)|^4\right\}$. $G_{\mathrm{mod}}$ varies depending on the input distribution, namely $G_{\mathrm{mod}}=2$ for a CSCG input $s(t)\sim\mathcal{CN}(0,1)$, $G_{\mathrm{mod}}=3$ for $s(t)\sim\mathcal{N}(0,1)$ and $G_{\mathrm{mod}}=l^2$ for the flash signaling distribution of \cite{Varasteh:2017b} characterized by $s=r e^{j\theta}$ with the phase $\theta$ uniformly distributed over $\left[0,2\pi\right)$ and the amplitude distributed according to the following probability mass function (with $l\geq 1$)
\begin{align}\label{flash_distr}
p_r(r)=\begin{cases}
1-\frac{1}{l^2}, \ & r=0,\\
\frac{1}{l^2}, \ & r=l.
\end{cases}
\end{align} 
\par Hence, transmit diversity with energy modulation brings a total gain in the fourth order term equal to the product of the diversity gain $G_{\mathrm{td}}$ and the modulation gain $G_{\mathrm{mod}}$.
  
\subsection{Transmit Diversity with Energy Waveform}\label{subsection_td_wf}
Instead of using an energy modulated single-carrier signal, we can use a deterministic in-phase multisine waveform with uniform power allocation on each antenna\footnote{Recall from \cite{Clerckx:2016b} that allocating power uniformly across all sinewaves leads to performance very close to the optimum in frequency flat channels.}. In such a case, \eqref{TD_mod_sys_mod} still holds but $s(t)$ is replaced by the baseband multisine waveform $s(t)=\sum_{n=0}^{N-1}\frac{1}{\sqrt{N}}e^{j n\Delta_{\mathrm{w}} t}$ with $\Delta_{\mathrm{w}}=2\pi\Delta_{\mathrm{f}}$ the inter-carrier frequency spacing. Assuming a frequency-flat channel, the received signal writes as in \eqref{TD_mod_sys_mod_rec}.
Following \cite{Clerckx:2016b}, we can then compute
\begin{align}
\bar{z}_{\mathrm{dc,td-wf}}=k_2 R_{\ant}\bar{P}_{\rf}+\frac{3}{2} k_4 R_{\ant}^2 \bar{P}_{\rf}^2 G_{\mathrm{td}}G_{\mathrm{wf}}
\end{align}
where $G_{\mathrm{wf}}=\frac{2N^2+1}{3N}\stackrel{N\nearrow}{\approx}\frac{2N}{3}$.
\par Hence, transmit diversity with energy waveform brings a total gain in the fourth order term equal to the product of the diversity gain $G_{\mathrm{td}}$ and the waveform gain $G_{\mathrm{wf}}$.

\section{Performance Analysis with Curve Fitting-based Nonlinear Rectenna Model}\label{nonlinear_model_fit_section}
The previous sections were based on a diode nonlinear model driven by the physics of the rectenna. We here study another rectenna nonlinear model that is based on fitting data using a polynomial. We use this model to assess the impact of fading on the harvested DC power and the potential gain of transmit diversity. Observations made in this section confirm those made in previous sections. Hence, despite the numerous differences between the physics-based model and the curve fitting-based model \cite{Clerckx:2018b,Clerckx:2018c}, both approaches lead to the same conclusion that fading and transmit diversity enhance the RF-to-DC conversion efficiency. This confirms that the benefits of fading and transmit diversity for WPT originate from the nonlinearity of the harvester and are not an artefact of the underlying nonlinear model.

\begin{figure}[t!]
\centering
\includegraphics[width=1.3 in,angle=-90]{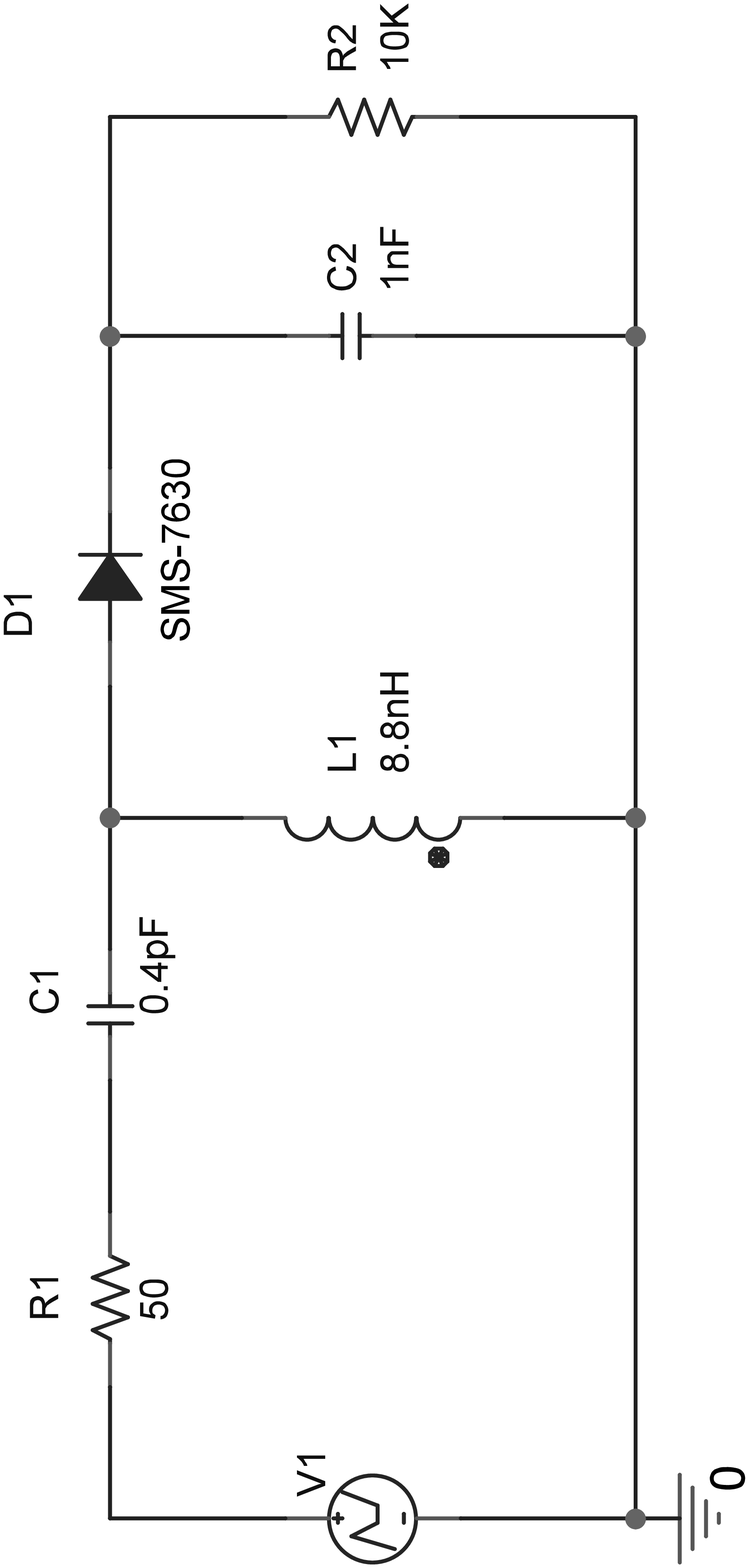}
\caption{Rectenna used for circuit simulations.}
\label{TD_schematic}
\vspace{-0.4cm}
\end{figure}

\subsection{A Curve Fitting-based Nonlinear Rectenna Model}\label{curve_fitting_subsec}

\par Since a curve fitting-based model requires actual data, we designed, optimized and simulated the rectenna circuit of Fig. \ref{TD_schematic}. We used a conventional single series rectifier circuit that consists of a rectifying diode, impedance matching circuit, and low pass filter. The Schottky diode Skyworks SMS7630 \cite{SMS7630} is chosen for the rectifying diode because it requires low biasing voltage level, which is suitable for low power rectifier. The impedance matching and low pass filter circuits are designed for an in-phase 4-tone ($N=4$) multisine input signal centered around 2.45GHz with an RF input power of -20dBm and with 2.5MHz inter-carrier frequency spacing. The load impedance R2 is chosen as $10\mathrm{K} \Omega$ in order to reach maximum RF-to-DC conversion efficiency with the 4-tone multisine waveform. The matching network capacitor C1, inductor L1 and output capacitor C2 values are optimized (using an iterative process) to maximize the output DC power under a given load impedance and for the given multisine input waveform at -20dBm RF input power. The chosen values are given by 0.4pF for C1, 8.8nH for L1, and 1nF for C2. The antenna impedance is set as $\mathrm{R1} = R_{\ant} = 50 \Omega$ and the voltage source $\mathrm{V1}$ is expressed as $\mathrm{V1} = v_{s}(t) = 2y(t) \sqrt{R_{\ant}}$.

\par Using PSpice circuit simulations, we evaluate in Fig. \ref{fitting_curve_model_CW} and Fig. \ref{fitting_curve_model_multisine} the DC power $P_{\dc}$ harvested at the load as a function of the RF input power $P_{\rf}$ for a continuous wave and multisine ($N=8$) input RF signals, respectively. Based on the collected data, we can then choose a model that fits the data. We fit the data (in a least-squares sense) using a polynomial as follows
\begin{equation}\label{EH_model_fit}
P_{\mathrm{dc,dB}}=a P_{\mathrm{rf,dB}}^2+b P_{\mathrm{rf,dB}}+c
\end{equation}
\vspace{-0.1cm}
with 
\vspace{-0.1cm}
\begin{align}
P_{\mathrm{dc,dB}}&=\log\left(P_{\mathrm{dc}}\right),\\
P_{\mathrm{rf,dB}}&=\log\left(P_{\mathrm{rf}}\right)=\log\left(|h|^2\right)+\bar{P}_{\mathrm{rf,dB}},\\
\bar{P}_{\mathrm{rf,dB}}&=\log\left(\bar{P}_{\rf}\right).
\end{align}
The coefficients $a$, $b$ and $c$ of the polynomials are given for two different RF input power ranges in the legends of Fig. \ref{fitting_curve_model_CW} and Fig. \ref{fitting_curve_model_multisine}, for continuous wave ($N=1$) and multisine ($N=8$) excitations, respectively. Note that, similarly to \cite{Alevizos:2018}, the fitting is performed in the log-log scale instead of using a linear scale. Using a sigmoid or a polynomial to fit data in the linear scale, as in \cite{Boshkovska:2015} and \cite{Xu:2017} respectively, may lead to severe inaccuracies in the low power regime as highlighted in \cite{Clerckx:2018c}. We note that a second degree polynomial fits the data relatively well, especially whenever fitted to the low input power regime\footnote{We are interested in the non-diode breakdown region, i.e. below -5dBm for the considered circuit, for the reasons discussed in Remark 5 of \cite{Clerckx:2018b}.} -40 dBm to -5 dBm, as per subfigures (b). Subfigures (c) illustrate the RF-to-DC conversion efficiency. Second degree fitting in the low power regime (-40 dBm to -5 dBm) appears relatively accurate, though RF-to-DC conversion efficiency is in general sensitive to any discrepancy in the curve fitting since it describes the ratio between $P_{\dc}$ and $P_{\rf}$. We also note that multisine excitation leads to higher harvested DC power and RF-to-DC conversion efficiency compared to continuous wave \cite{Trotter:2009,Clerckx:2016b,Clerckx:2017}.

\vspace{-0.2cm}
\subsection{Effect of Fading on Harvested Energy}\label{fit_model_fading}
Similarly to Section \ref{Fading_section}, we here assess the impact of fading on harvested DC power. We assume that the fading coefficient $h$ is modeled as CSCG with $\mathbb{E}\left[|h|^2\right]=1$. With a fading channel, $P_{\rf}$ fluctuates as a function of $h$ around its mean $\bar{P}_{\rf}$. As a consequence, $P_{\dc}$ also fluctuates as a function of $h$. For a given realization of $h$, $P_{\rf}$ relates to $P_{\dc}$ through \eqref{EH_model_fit}. Defining the average DC power as $\bar{P}_{\dc}=\mathbb{E}_{h}\left[P_{\dc}\right]$, with the expectation taken over the channel distribution, and making use of the nonlinear model \eqref{EH_model_fit} with $P_{\mathrm{rf}}=|h|^2 \bar{P}_{\rf}$, we write 
\begin{align}
\bar{P}_{\dc}&=\mathbb{E}_{h}\left[e^{P_{\mathrm{dc,dB}}}\right]=\mathbb{E}_{h}\left[e^{a \left(\log(|h|^2 \bar{P}_{\rf})\right)^2+b \log(|h|^2 \bar{P}_{\rf})+c}\right],\nonumber\\
&=P_{\dc,\mathrm{no\;fading}} \;e_{\mathrm{fading}} \label{average_Pdc_fit}
\end{align}
where
\begin{align}
P_{\dc,\mathrm{no\;fading}}&=e^{a \bar{P}_{\mathrm{rf,dB}}^2+b \bar{P}_{\mathrm{rf,dB}}+c},\\
e_{\mathrm{fading}}&=\mathbb{E}_{h}\left[e^{a \left(\log(|h|^2)\right)^2+(2 a \bar{P}_{\mathrm{rf,dB}}+ b)\log(|h|^2) }\right].\label{G_fading}
\end{align}
According to \eqref{average_Pdc_fit}, the average harvested DC power $\bar{P}_{\dc}$ with fading can be written as the product of the harvested DC power $P_{\dc,\mathrm{no\;fading}}$ without fading and a fading gain $e_{\mathrm{fading}}$ that characterizes the impact of channel fading on the harvested DC power. Interestingly, \eqref{average_Pdc_fit} can be equivalently written as
\begin{align}
\bar{P}_{\dc}&=\overbrace{\bar{P}_{\rf} \;e_{\mathrm{rf-dc}}}^{P_{\dc,\mathrm{no\;fading}}} \;e_{\mathrm{fading}},\nonumber\\
&=\bar{P}_{\rf} \;\underbrace{e_{\mathrm{rf-dc}} \;e_{\mathrm{fading}}}_{\bar{e}_{\mathrm{rf-dc}}}, \label{average_Pdc_fit_2}
\end{align}
where $e_{\mathrm{rf-dc}}=P_{\dc,\mathrm{no\;fading}}/\bar{P}_{\rf}$ is the conventional RF-to-DC conversion efficiency (with e.g. continuous wave or multisine excitation) in the absence of fading and $\bar{e}_{\mathrm{rf-dc}}=\bar{P}_{\dc}/\bar{P}_{\rf}$ is the RF-to-DC conversion efficiency in the presence of fading. Interestingly, the RF-to-DC conversion efficiency in the presence of fading $\bar{e}_{\mathrm{rf-dc}}$ can be split into two contributions: the conventional RF-to-DC conversion efficiency in the absence of fading $e_{\mathrm{rf-dc}}$ and a fading gain $e_{\mathrm{fading}}$ that accounts for the impact of fading on the harvester nonlinearity. 

\begin{figure}[t!]
 \centering
\begin{subfigmatrix}{3}
\subfigure[Curve fitting for input RF power range -40dBm to 20dBm. Second degree polynomial: $a=-0.0977$, $b=-0.9151$, $c=-11.1684$. First degree polynomial character: $a=0$, $b=0.8848$, $c=-4.6926$.]{\label{a}\includegraphics[width=3.3 in]{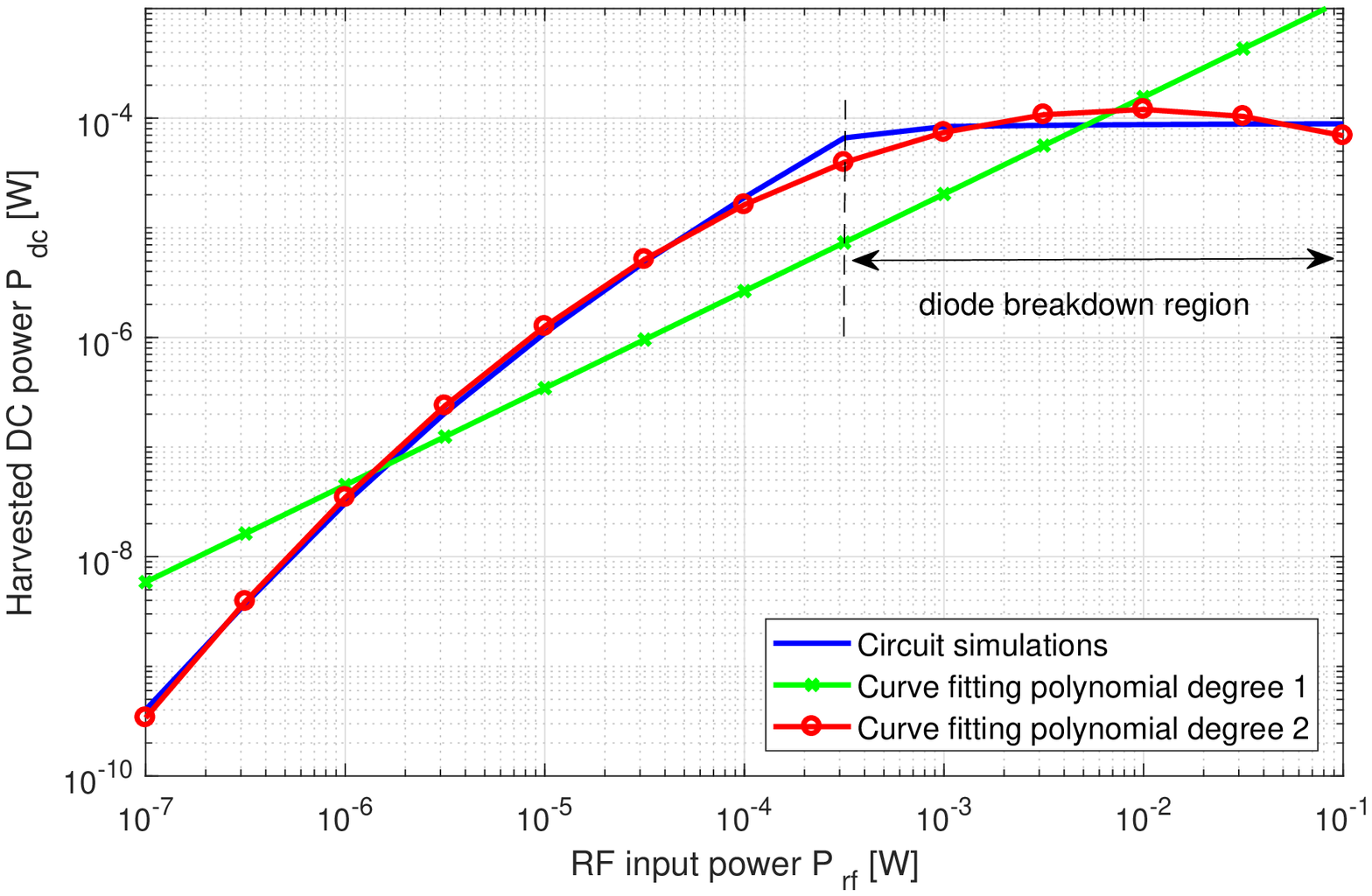}}
\subfigure[Curve fitting for input RF power range -40dBm to -5dBm. Second degree polynomial: $a=-0.0669$, $b=-0.1317$, $c=-6.3801$. First degree polynomial: $a=0$, $b=1.4848$, $c=2.9250$.]{\label{b}\includegraphics[width=3.3 in]{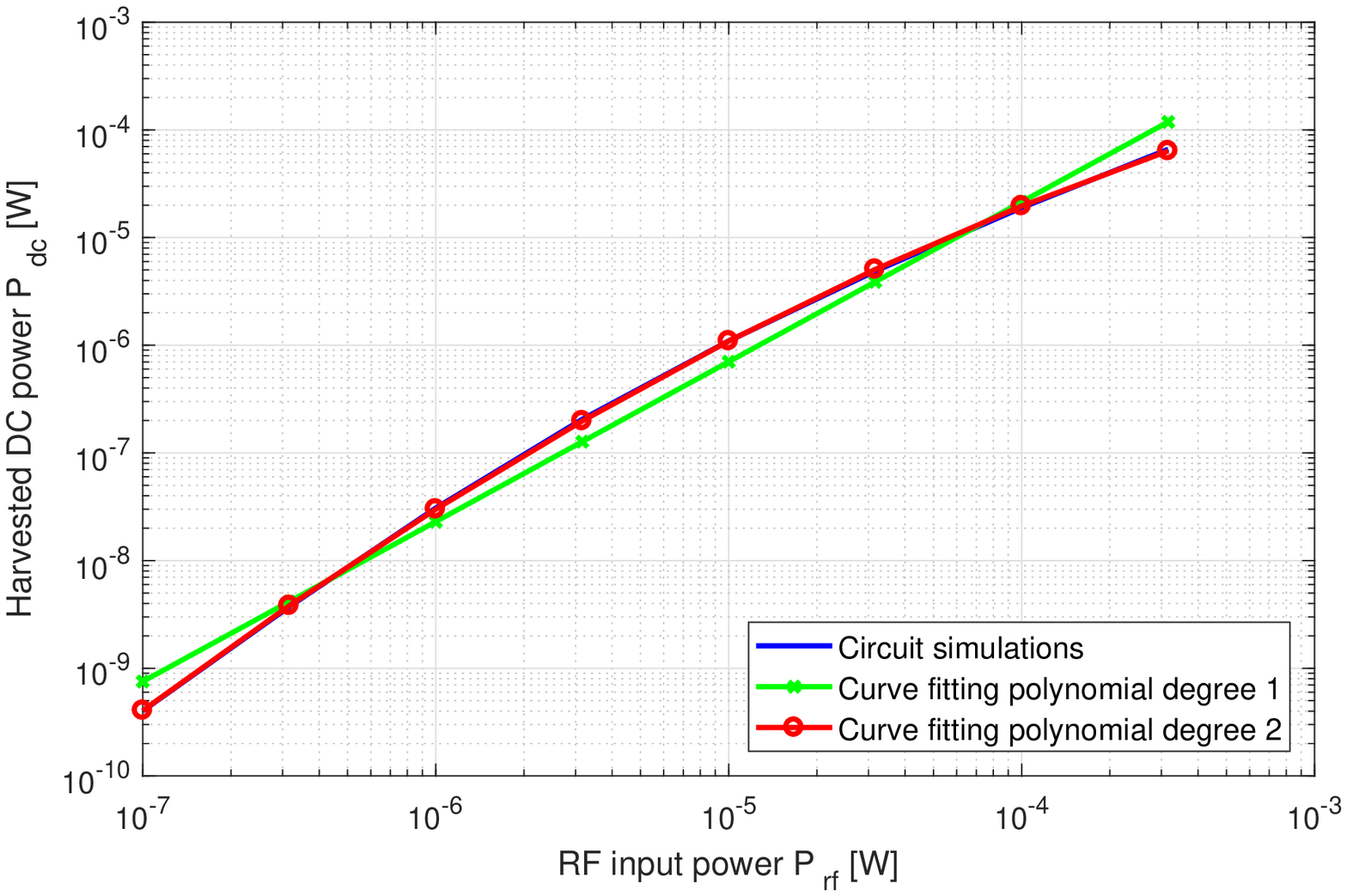}}
\subfigure[Corresponding RF-to-DC conversion efficiency $P_{\dc}/P_{\rf}$. ]{\label{c}\includegraphics[width=3.3 in]{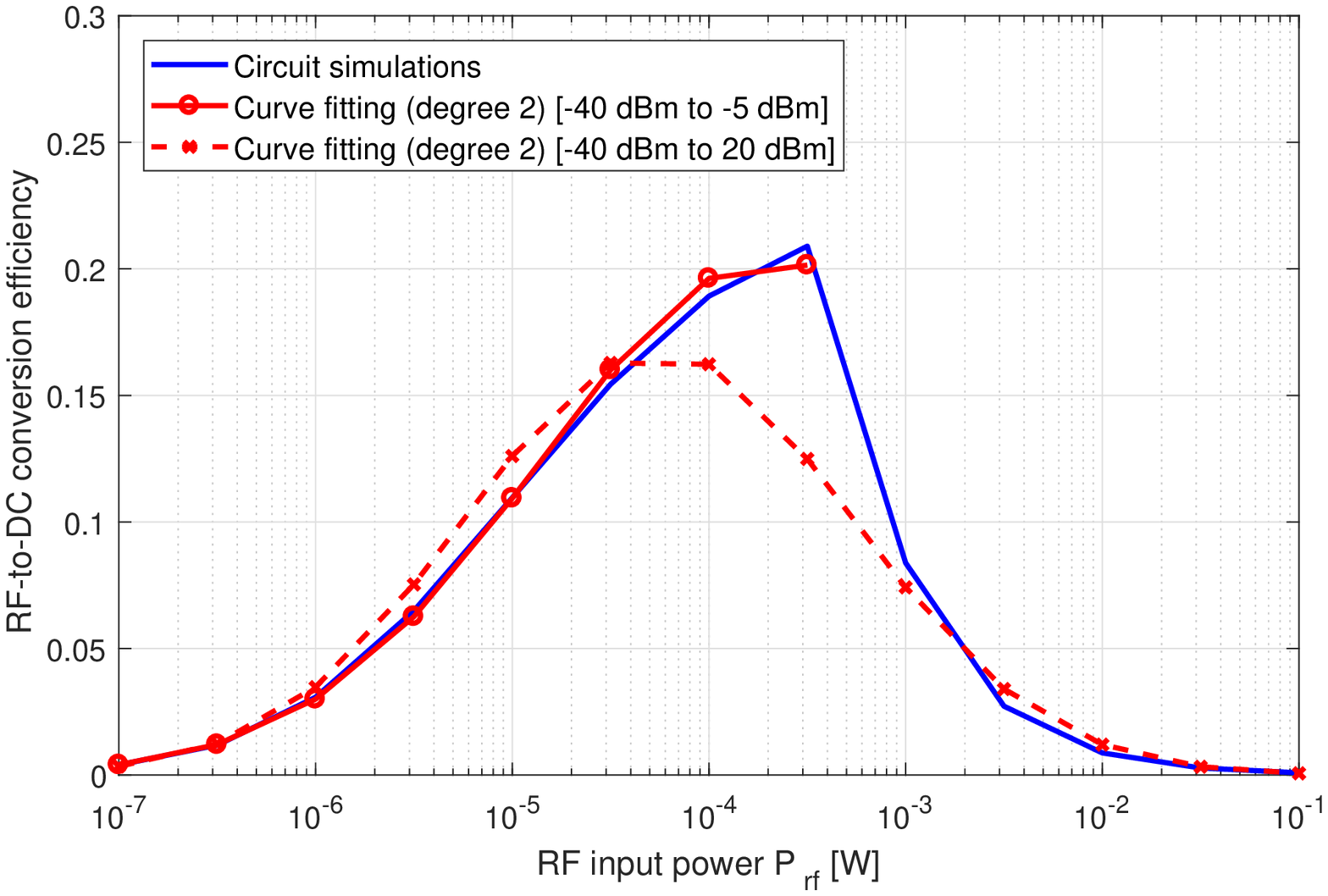}}
\end{subfigmatrix}
 \caption{Circuit evaluations of harvested DC power $P_{\dc}$ versus RF input power $P_{\rf}$ for a \textit{continuous wave} ($N=1$) excitation and corresponding curve fitting model for the rectenna in Fig. \ref{TD_schematic}.} \label{fitting_curve_model_CW}
\vspace{-0.3cm}
\end{figure}

\begin{figure}[t!]
 \centering
\begin{subfigmatrix}{3}
\subfigure[Curve fitting for input RF power range -40dBm to 20dBm. Second degree polynomial: $a=-0.1089$, $b=-1.1660$, $c=-12.0041$. First degree polynomial: $a=0$, $b=0.8396$, $c=-4.7881$.]{\label{a}\includegraphics[width=3.3 in]{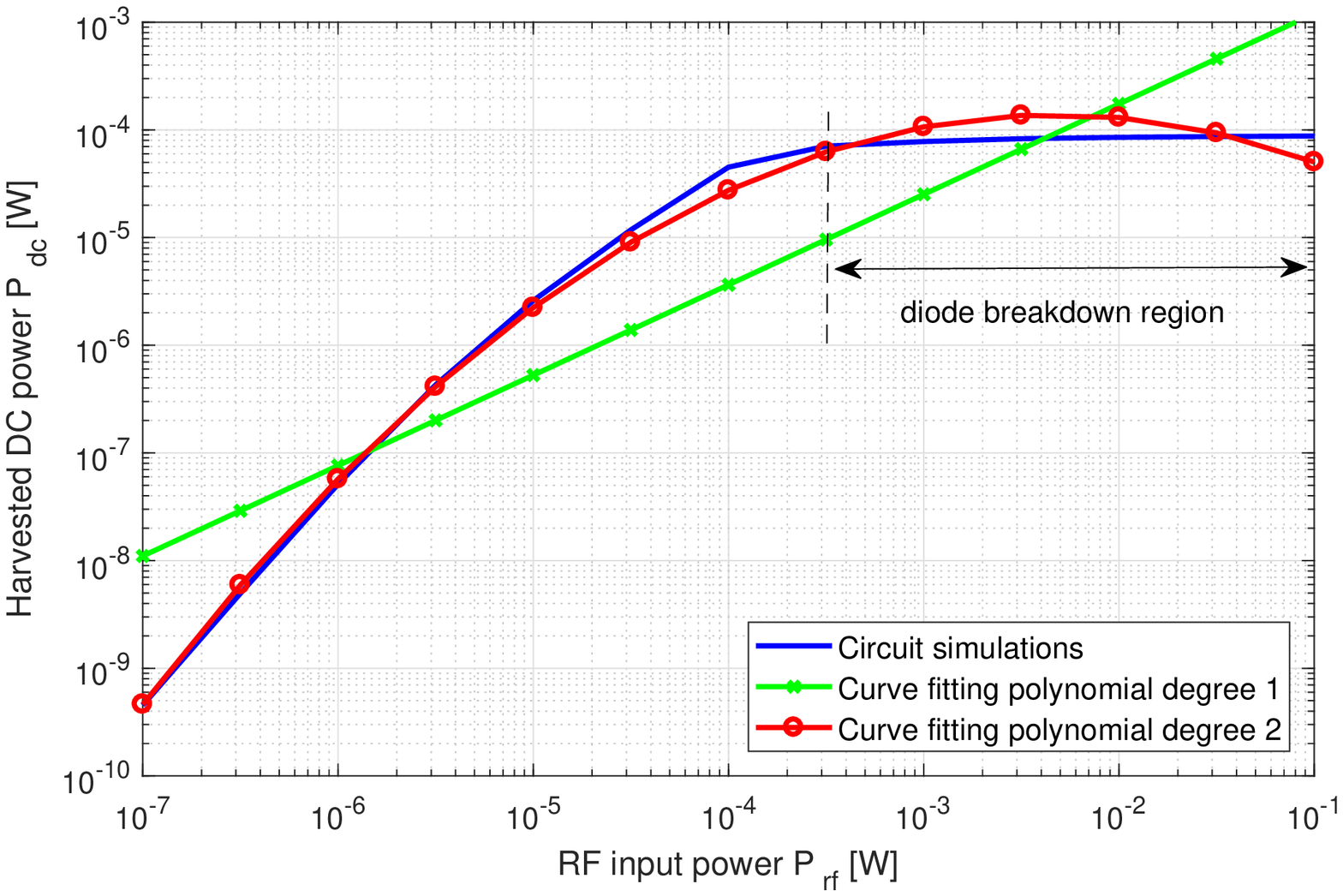}}   
\subfigure[Curve fitting for input RF power range -40dBm to -5dBm. Second degree polynomial: $a=-0.1105$, $b=-1.1468$, $c=-11.4342$. First degree polynomial: $a=0$, $b=1.5239$, $c=3.9400$.]{\label{b}\includegraphics[width=3.3 in]{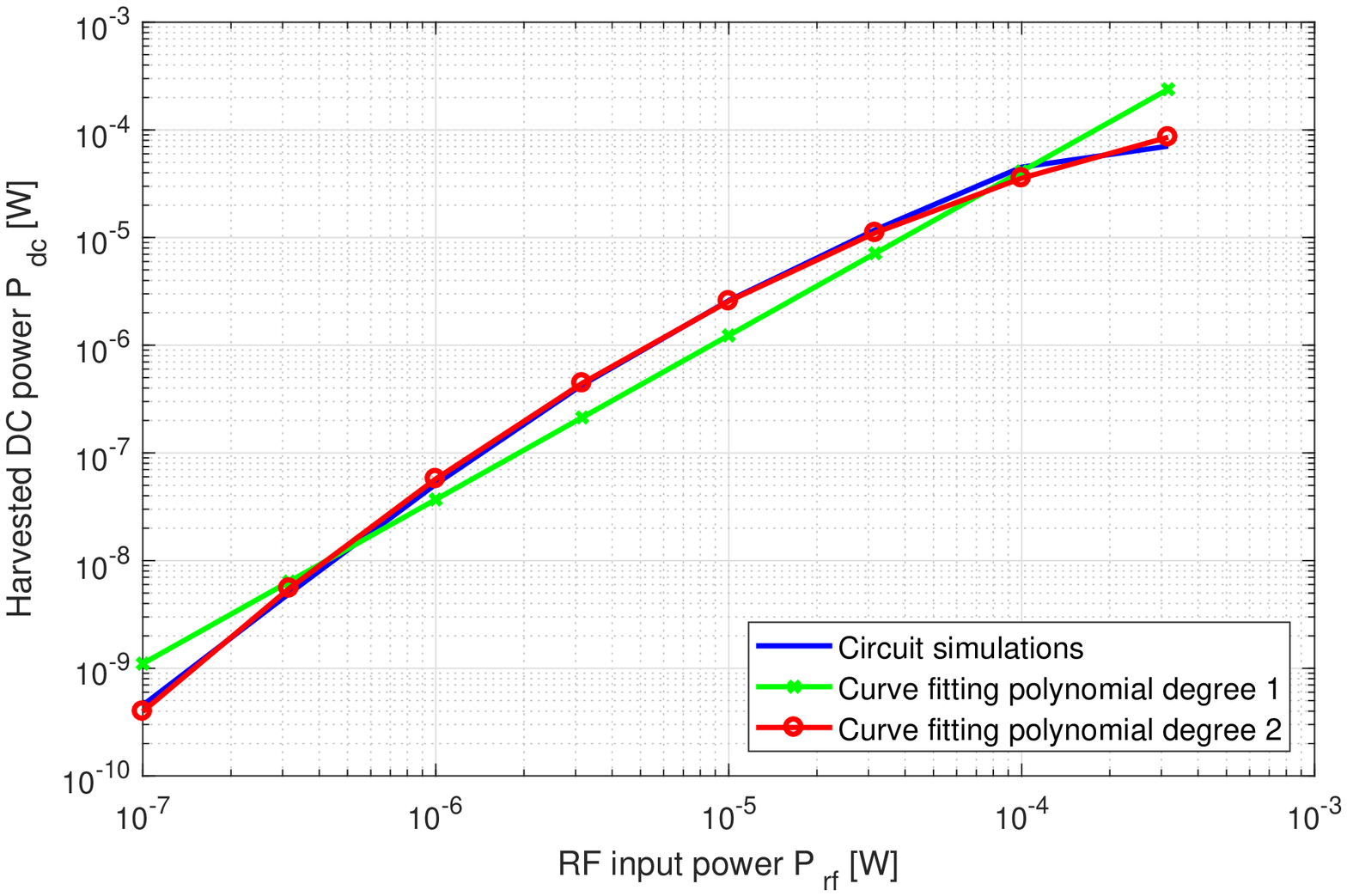}}
\subfigure[Corresponding RF-to-DC conversion efficiency $P_{\dc}/P_{\rf}$. ]{\label{c}\includegraphics[width=3.3 in]{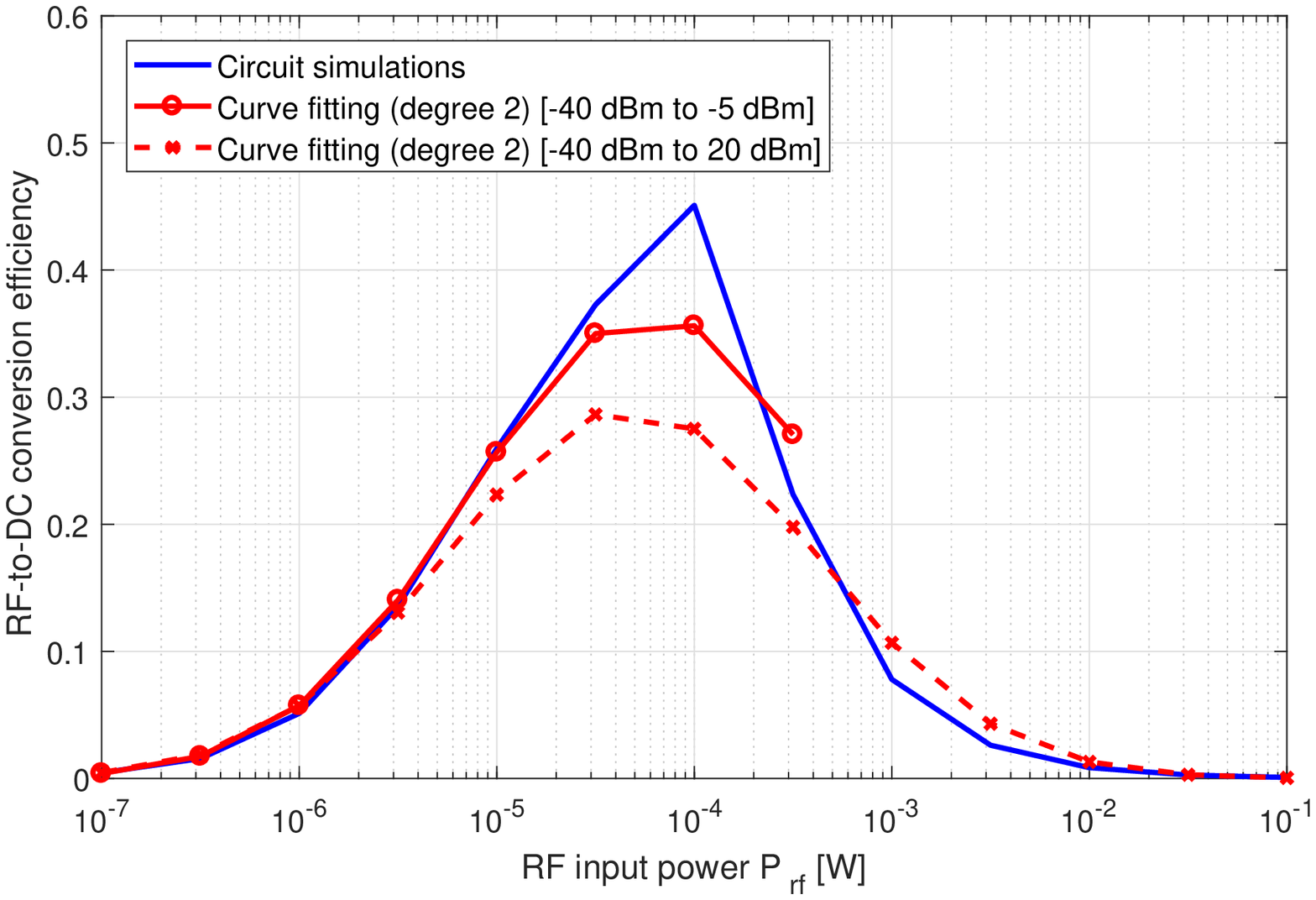}}
\end{subfigmatrix}
 \caption{Circuit evaluations of harvested DC power $P_{\dc}$ versus RF input power $P_{\rf}$ for a \textit{multisine} ($N=8$) excitation and corresponding curve fitting model for the rectenna in Fig. \ref{TD_schematic}.} \label{fitting_curve_model_multisine}
\vspace{-0.3cm}
\end{figure}

\par
\begin{figure} [t!]
\centering
\includegraphics[width=3.3 in]{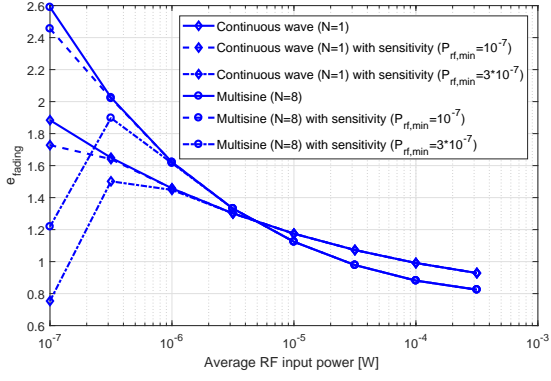}
\caption{\textit{Modeled} fading gain $e_{\mathrm{fading}}$ (with and without rectifier sensitivity) with a continuous wave ($N=1$) and multisine ($N=8$) excitation for an average RF input power $\bar{P}_{\rf}$ ranging from -40dBm to -5dBm. Second degree polynomial fitting parameters taken from Fig. \ref{fitting_curve_model_CW}(b), \ref{fitting_curve_model_multisine}(b).}
\label{e_fading_figure}
\end{figure}

\begin{figure} 
\centering
\includegraphics[width=3.3 in]{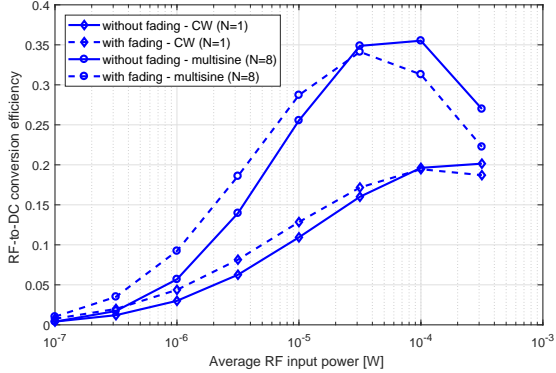}
\caption{\textit{Modeled} RF-to-DC conversion efficiency without fading $e_{\mathrm{rf-dc}}$ and with fading $\bar{e}_{\mathrm{rf-dc}}$ for a continuous wave ($N=1$) and multisine ($N=8$) excitation in the average RF input power $\bar{P}_{\rf}$ range of -40dBm to -5dBm. Second degree polynomial fitting parameters taken from Fig. \ref{fitting_curve_model_CW}(b), \ref{fitting_curve_model_multisine}(b).}
\label{RF_DC_conversion_fading_fit}
\end{figure}

 Defining $d=2 a \bar{P}_{\mathrm{rf,dB}}+ b$ and $X=|h|^2$, and noting that $X\sim \mathrm{EXPO}(1)$, the fading gain $e_{\mathrm{fading}}$ can be expressed as
\begin{align}
e_{\mathrm{fading}}&=\mathbb{E}_{X}\left[e^{a \left(\log(X)\right)^2+d \log(X)}\right],\nonumber\\
&=\mathbb{E}_{X}\left[X^{d+a \log(X)}\right],\nonumber\\
&=\int_{0}^{\infty}x^{d+a \log(x)} e^{-x} dx.\label{e_fading}
\end{align}
Numerical integration can be used to compute the last integral. The fading gain $e_{\mathrm{fading}}$ is illustrated in Fig. \ref{e_fading_figure} for continuous wave ('Continuous wave (N=1)') and multisine ('Multisine (N=8)') excitations using the fitting parameters taken from Fig. \ref{fitting_curve_model_CW} and Fig. \ref{fitting_curve_model_multisine}, respectively. It appears that $e_{\mathrm{fading}}$ is larger than one for most input power, especially in the low input power regime, suggesting that fading is beneficial to energy harvesting. This confirms observations made in Section \ref{Fading_section}. The gain can be explained again by the convexity of the $P_{\dc}$-$P_{\rf}$ relationship in the low power regime. The fading gain leads to an increase of the RF-to-DC conversion efficiency in the low input power regime as illustrated in Fig. \ref{RF_DC_conversion_fading_fit}. We also note in Fig. \ref{e_fading_figure} that the fading gain with a multisine is smaller than with a continuous wave for high RF input power levels and larger for low RF input power levels. The behavior is again observed because of the convexity at low input power and concavity at high input powers of the $P_{\dc}$-$P_{\rf}$ relationship. As it appears from Fig. \ref{RF_DC_conversion_fading_fit}, $e_{\mathrm{rf-dc}}$ without fading increases more rapidly with a multisine than with a continuous wave at low input power, but decreases more rapidly at high input power levels.

\begin{remark}\label{remark_sensitivity} Circuit simulations have not been conducted below -40dBm because such a power level is likely too low to be harvested due to the limited sensitivity of the rectifier. Indeed, we note in Fig. \ref{fitting_curve_model_CW}(c) and \ref{fitting_curve_model_multisine}(c) that the RF-to-DC conversion efficiency is close to 0 at around -40dBm. Though the polynomial-based fitting \eqref{EH_model_fit} was done using the available data in the range -40dBm to 20dBm, we assumed that the same polynomial can be used for power levels below -40dBm to compute the fading gain in \eqref{e_fading} (and transmit diversity gain in \eqref{e_td}). As in \cite{Alevizos:2018,Alevizos:2018a}, we can improve the analysis by accounting for the limited sensitivity of the rectifier in the very low power regime by stating
\begin{align}\label{sensitivity}
P_{\mathrm{dc,dB}}=\begin{cases}
\eqref{EH_model_fit}, \ & P_{\mathrm{rf}}\geq P_{\mathrm{rf,min}},\\
0, \ & \mathrm{otherwise},
\end{cases}
\end{align}
where $P_{\mathrm{rf,min}}$ refers to the sensitivity threshold below which the rectifier does not turn on. Doing so, the fading gain in \eqref{e_fading} can be adjusted as follows
\begin{equation}
e_{\mathrm{fading}}=\int_{x_{\mathrm{min}}}^{\infty}x^{d+a \log(x)} e^{-x} dx,\label{e_fading_sensitivity}
\end{equation}
where $x_{\mathrm{min}}=P_{\mathrm{rf,min}}/\bar{P}_{\mathrm{rf}}$. As illustrated in Fig. \ref{e_fading_figure}, when the rectifier sensitivity is taken into account, the fading gain decreases in the very low power regime as $P_{\mathrm{rf,min}}$ increases.
\end{remark}

\begin{figure} [t!]
\centering
\includegraphics[width=3.3 in]{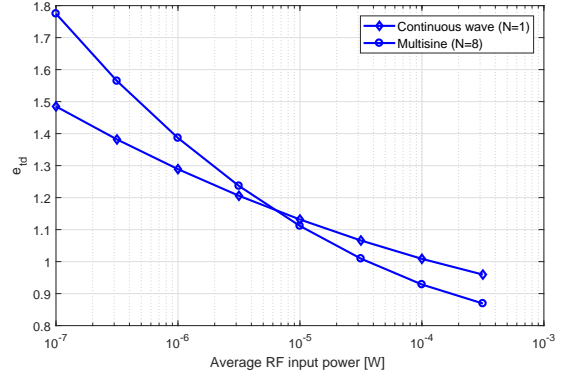}
\caption{\textit{Modeled} transmit diversity gain $e_{\mathrm{td}}$ with a continuous wave ($N=1$) and multisine ($N=8$) excitation for an average RF input power $\bar{P}_{\rf}$ ranging from -40dBm to -5dBm. Second degree polynomial fitting parameters taken from Fig. \ref{fitting_curve_model_CW}(b), \ref{fitting_curve_model_multisine}(b).}
\label{e_diversity_figure}
\end{figure}

\begin{figure} 
\centering
\includegraphics[width=3.3 in]{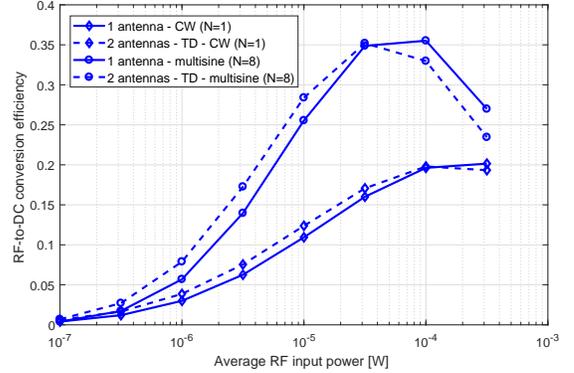}
\caption{\textit{Modeled} RF-to-DC conversion efficiency with one transmit antenna without fading $e_{\mathrm{rf-dc}}$ and with two antenna-transmit diversity $e_{\mathrm{rf-dc,td}}$ for a continuous wave ($N=1$) and multisine ($N=8$) in the average RF input power $\bar{P}_{\rf}$ range of -40dBm to -5dBm. Second degree polynomial fitting parameters taken from Fig. \ref{fitting_curve_model_CW}(b), \ref{fitting_curve_model_multisine}(b).}
\label{RF_DC_conversion_TD_fit}
\end{figure}

\subsection{Effect of Transmit Diversity on Harvested Energy}\label{fit_model_TD_analysis}

Similarly to Subsection \ref{basic_TD_section_sweeping}, we here assess the benefits of transmit diversity on the harvested DC power with the curve fitting-based nonlinear model. Let us consider the simplest case of two transmit antennas ($M=2$) with $h_m=1$ and the phases $\psi_1$ and $\psi_2$ independent and uniformly distributed over $2\pi$. The effective channel can be written as $h(t)=e^{j \psi_1(t)}+e^{j \psi_2(t)}$ and $\left|h(t)\right|^2=2\left(1+\cos\left(\psi_1(t)-\psi_2(t)\right)\right)$. Defining the random variable $U=\psi_1(t)-\psi_2(t)$, also uniformly distributed over $2\pi$, and equally splitting the transmit power $P$ among the two antennas ($P/2$ on each antenna), the effect of transmit diversity on the harvested DC power can be computed following the same steps as in Subsection \ref{fit_model_fading} by replacing $|h|^2$ with $1+\cos\left(U\right)$. Following \eqref{average_Pdc_fit_2}, we therefore write 
\begin{align}
\bar{P}_{\dc,\mathrm{td}}&=\overbrace{\bar{P}_{\rf} \;e_{\mathrm{rf-dc}}}^{P_{\dc,\mathrm{no\;fading}}} \;e_{\mathrm{td}},\nonumber\\
&=\bar{P}_{\rf} \;\underbrace{e_{\mathrm{rf-dc}} \;e_{\mathrm{td}}}_{e_{\mathrm{rf-dc,td}}} \label{average_Pdc_fit_td}
\end{align}
where the transmit diversity gain $e_{\mathrm{td}}$ is obtained by replacing $X$ in \eqref{e_fading} by $1+\cos\left(U\right)$, such that
\begin{align}
e_{\mathrm{td}}&=\mathbb{E}_{U}\left[e^{a \left(\log(1+\cos\left(U\right))\right)^2+d \log(1+\cos\left(U\right))}\right],\nonumber\\
&=\mathbb{E}_{U}\left[\left(1+\cos\left(U\right)\right)^{d+a \log(1+\cos\left(U\right))}\right],\nonumber\\
&=\int_{0}^{2\pi}\left(1+\cos\left(u\right)\right)^{d+a \log(1+\cos\left(u\right))} \frac{1}{2\pi} du. \label{e_td}
\end{align}
Numerical integration can be used to compute the last integral. The transmit diversity gain $e_{\mathrm{td}}$ is illustrated in Fig. \ref{e_diversity_figure} for continuous wave and multisine excitations using the fitting parameters taken from Fig. \ref{fitting_curve_model_CW} and Fig. \ref{fitting_curve_model_multisine}, respectively. It appears that $e_{\mathrm{td}}$ is larger than one for most input power, especially in the low input power regime, suggesting that transmit diversity is beneficial to energy harvesting. This is also inline with observations made in Subsection \ref{basic_TD_section_sweeping} and Section \ref{section_TD_mod_wf}. The transmit diversity gain $e_{\mathrm{td}}$ leads to an RF-to-DC conversion efficiency with transmit diversity $e_{\mathrm{rf-dc,td}}$ larger than the RF-to-DC conversion efficiency achieved with a single transmit antenna $e_{\mathrm{rf-dc}}$, as illustrated in Fig. \ref{RF_DC_conversion_TD_fit}. Following Remark \ref{remark_sensitivity}, accounting for the limited rectifier sensitivity would lead to a decrease of $e_{\mathrm{td}}$ in the very low power regime. 

\section{Performance Evaluations}\label{simulations}

Previous sections highlighted the benefit of transmit diversity in WPT with both continuous wave or more complicated modulation/waveform. The gains were predicted from the analysis using two different nonlinear models of the energy harvester. We here evaluate the performance benefits of transmit diversity using circuit simulations and real-time over-the-air experimentation, and contrast the results with the analysis.

\subsection{Circuit Simulations}

\begin{figure} [t!]
\centering
\includegraphics[width=3.3 in]{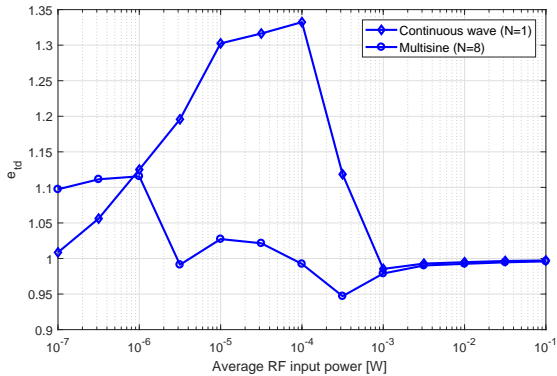}
\caption{\textit{Simulated} transmit diversity gain $e_{\mathrm{td}}$ with a continuous wave ($N=1$) and multisine ($N=8$) for an average RF input power $\bar{P}_{\rf}$ ranging from -40dBm to 20dBm.}
\label{e_diversity_figure_simulations}
\end{figure}

\begin{figure} [t!]
\centering
\includegraphics[width=3.3 in]{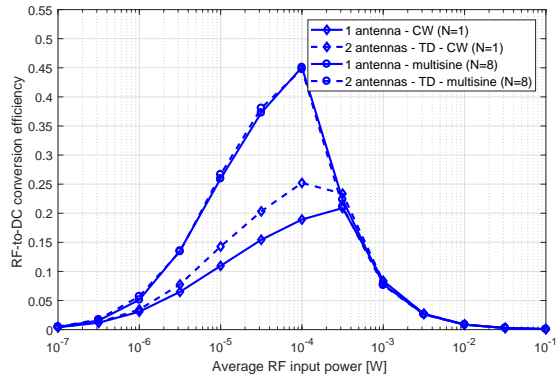}
\caption{\textit{Simulated} RF-to-DC conversion efficiency with one transmit antenna without fading $e_{\mathrm{rf-dc}}$ and with two antenna-transmit diversity $e_{\mathrm{rf-dc,td}}$ for a continuous wave ($N=1$) and multisine ($N=8$) in the average RF input power $\bar{P}_{\rf}$ range of -40dBm to 20dBm.}
\label{RF_DC_conversion_TD_simulations}
\end{figure}

\par We consider a two-antenna ($M=2$) transmit diversity system over a deterministic channel with $h_m=1$ (similarly to the analysis). Using Matlab, we generate the input voltage source V1 to the realistic rectenna of Fig. \ref{TD_schematic}. This is done by generating the transmit signal $s(t)$, using either a continuous wave (CW) or multisine ($N=8$), passing it through the two-antenna transmit diversity strategy whose phase is randomly swept at a 2.5 MHz rate, and converting the received signal $y(t)$ (assuming $h_m=1$) to the input voltage source V1. We then feed the input voltage source V1 into the PSpice circuit simulator. The duration of the generated signal was determined to be long enough (about 70 $\mathrm{\mu s}$) for the rectifier to be in a steady state response mode. For each input power level ranging from -40dBm to 20dBm, we repeat the same process for more than 150 times in order to average out over a sufficient number of realizations of the phase. 
\par We chose a transmit diversity phase change rate of 2.5 MHz though we also considered several phase change rates between 1 MHz and 10 MHz. The periods of those phase changes range from 0.1 $\mathrm{\mu s}$ to 1 $\mathrm{\mu s}$. As stated in Subsection \ref{curve_fitting_subsec}, the rectenna has been optimized for a 4-tone multisine input waveform with an inter-carrier frequency spacing $\Delta_{\mathrm{f}}=2.5$ MHz. Such a waveform generates large peaks with a periodicity of $1/\Delta_{\mathrm{f}}=0.4$ $\mathrm{\mu s}$. The RC constant of the low pass filter is $10 \mathrm{K}\Omega \times 1 \mathrm{nF} = 10 \mathrm{\mu s}$. The period of the phase change is therefore smaller than the rectenna's RC constant. We did not observe significant performance gap between different phase change rates. We set the phase change rate to 2.5 MHz as mentioned above because it is applicable for the simulations and the prototype.

\par Fig. \ref{e_diversity_figure_simulations} illustrates the simulated gain of transmit diversity with two transmit antennas ($M=2$). Hence, in Fig. \ref{e_diversity_figure} and Fig. \ref{e_diversity_figure_simulations}, $e_{\mathrm{td}}$ refers to the transmit diversity gain obtained from analysis/modeling and simulations, respectively. Fig. \ref{RF_DC_conversion_TD_simulations} displays the simulated RF-to-DC conversion efficiency (in contrast to the modeled one in Fig. \ref{RF_DC_conversion_TD_fit}) with and without transmit diversity. A performance gain of transmit diversity with a \textit{continuous wave} input is observed in circuit simulations, therefore validating the prediction from the analysis. Though a transmit diversity gain with a continuous wave is observed in both analysis (Fig. \ref{e_diversity_figure}) and simulations (Fig. \ref{e_diversity_figure_simulations}), the general behavior is not the same in both figures. Indeed, while the analysis highlights a decreasing gain as the input power level increases, the simulations display larger gains in the practical RF input power level range of -20dBm to -10dBm. Recall that the rectenna of Fig. \ref{TD_schematic} was designed for an RF input power of -20dBm and that the limited sensitivity of the rectifier was not modeled in the very low power regime in Fig. \ref{e_diversity_figure}, which partly explain why the simulated transmit diversity gain is higher (than the modeled one) at -20dBm RF input power level and lower at -40dBm. The circuit simulations suggest that the limited rectifier sensitivity is indeed observed at very low power levels since the simulated transmit diversity gain displays a severe drop similar to the one illustrated in Fig. \ref{e_fading_figure} (for $P_{\mathrm{rf,min}}=3e^{-7}$). On the other hand, transmit diversity appears less beneficial when used in conjunction with \textit{multisine} waveform. Recall from the analysis in subsection \ref{subsection_td_wf} and from Figs. \ref{e_diversity_figure} and \ref{RF_DC_conversion_TD_fit} that transmit diversity was expected to provide an additional gain on top of the multisine waveform gain. It seems from circuit simulations that this additional gain is rather negligible. This is partially due to the rectifier design. Since the rectifier used in the simulations has been optimized for a 4-tone input signal, its has a limited ability to exploit the large fluctuations induced by transmit diversity with an 8-tone waveform. If we could adjust the rectifier (and its load) as a function of the input signal, we could see more gain of combining multisine and transmit diversity. Nevertheless, it is somewhat intuitive that, in practice, we observe a diminishing return of combining transmit diversity and multisine waveform. Indeed, the multisine waveform already significantly boosts the RF-to-DC conversion efficiency as evidenced by Fig. \ref{RF_DC_conversion_TD_simulations}, therefore leaving less room for improvement for transmit diversity. 

\vspace{-0.2cm}
\subsection{Prototyping and Experimentation}

\begin{figure}
\centering
\includegraphics[width=2.5 in]{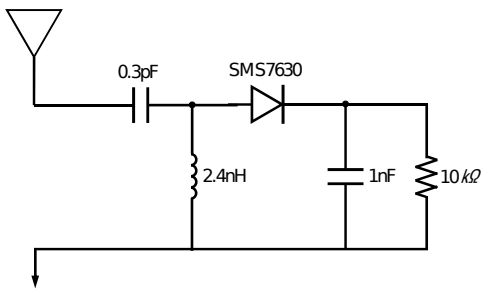}
\caption{Rectenna used in the experiment.}
\label{proto_rectenna}
\end{figure}

Leveraging the early WPT prototype reported in \cite{Kim:2017}, we have implemented a two-antenna transmit diversity prototype to verify that transmit diversity provides in real world conditions the performance benefits predicted from the analysis and simulation results. The prototype is divided into two parts, Tx and Rx, respectively. Rx part is a simple rectenna. The rectifier printed circuit board (PCB) is fabricated with a $\lambda /4$ length of microstrip, 1005 (1.0 mm x 0.5 mm lumped element package size) passive elements pads for $\pi$-matching network, and followed by a Schottky diode rectifier circuit. The same diode and low pass filter circuit as the rectenna used for the circuit simulations in Fig. \ref{TD_schematic} are implemented in the prototype. The values of the matching network components are modified to fit the real fabricated PCB. We use universal 2.4 GHz band WiFi antennas with 3dBi gain. Fig. \ref{proto_rectenna} illustrates the circuit diagram of the prototype rectenna used in the experiment. 

\par
\begin{figure}
\centering
\includegraphics[width=3.1 in]{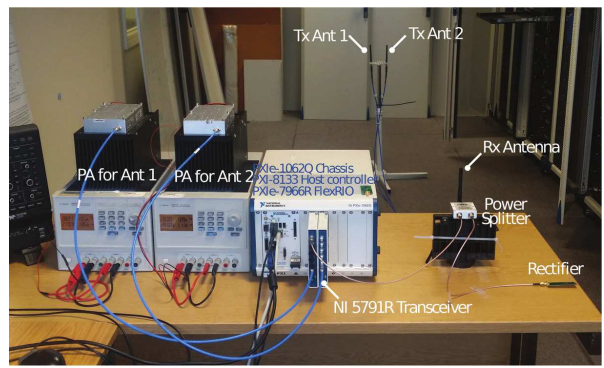}
\caption{Experimental setup with the SDR (PXIe-7966R and NI 5791R) implementing the transmit diversity strategy, the two PAs linked to Tx antenna 1 and 2, and the Rx antenna connected to the rectifier.}
\label{MISO_equipment}
\end{figure}

 Tx part is an RF signal generator with 2.45GHz center frequency implemented using a software defined radio (SDR) relying on NI FlexRIO PXIe-7966R and NI 5791R transceiver module. Tx signal generator continuously generates the selected signal type such as CW (continuous wave), multisine ($N=8$), TD-CW (transmit diversity with CW), and TD-multisine (transmit diversity with multisine). The experiment is carried out at two different locations. The Rx antenna locations are fixed and Tx antennas are moved to the test locations. Both locations exhibit line-of-sight (LoS) between Tx and Rx, but different Tx-Rx distances of 2.5m and 5m. The received DC voltage measurement is carried out five times, and each measurement, non-continuous in time, is one minute long. Final results are averaged value of 5 minutes total for each case. Fig. \ref{MISO_equipment} is a picture of the overall experimental setup.

\par
\begin{figure}
\centering
\includegraphics[width=3.3 in]{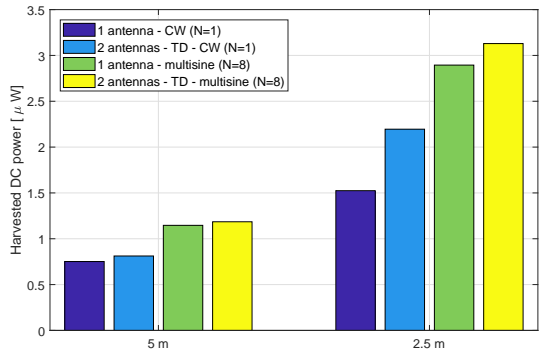}
\caption{\textit{Measured} harvested DC power with a continuous wave (CW), transmit diversity ($M=2$) with continuous wave, multisine ($N=8$) and transmit diversity with multisine ($N=8$).}
\label{Fig_TD_measurement_DC_power}
\end{figure}

\begin{figure}
\centering
\includegraphics[width=3.3 in]{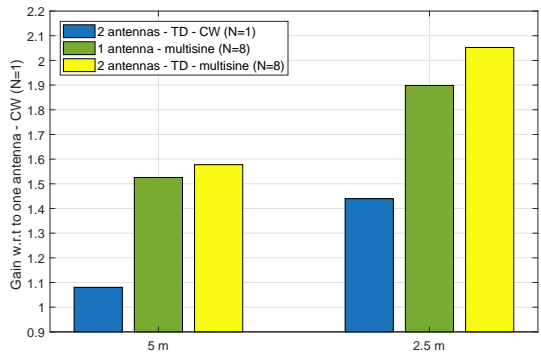}
\caption{\textit{Measured} relative gain (in terms of harvested DC power) with transmit diversity ($M=2$) with continuous wave, multisine ($N=8$) and transmit diversity with multisine ($N=8$) over a continuous wave (CW).}
\label{Fig_TD_measurement_gain}
\vspace{-0.2cm}
\end{figure}

 Fig. \ref{Fig_TD_measurement_DC_power} illustrates the harvested DC power measured at the load of the rectifier for various transmit strategies, namely a single antenna with continuous wave (CW), a two-antenna transmit diversity (TD) with continuous wave, a single antenna multisine ($N=8$) and a two-antenna transmit diversity with multisine ($N=8$). Two different distances of 2.5 m and 5 m from the transmitter are considered, corresponding to a relatively high and low input power to the energy harvester, respectively. We observe that measurements are inline with theory and circuit simulations, with transmit diversity providing gain with both continuous wave and multisine waveforms, though larger with continuous wave and at a moderate distance of 2.5 m. In Fig. \ref{Fig_TD_measurement_gain}, we plot the same results but in a different way to highlight the relative gain over the DC power harvested with the single antenna continuous wave. TD with CW offers a gain over CW of about 8\% and 45\%, at 5 m and 2.5 m, respectively. By combining TD with multisine, a combined transmit diversity and multisine ($N=8$) waveform gain is observed of about 58\% and 105\% over a single antenna CW. Recall that those gains are obtained without any need for CSIT and that, in a multi-user deployment, each device would benefit from this performance enhancement, irrespectively of the number of devices. It is also worth reminding that the prototype relies on two transmit antennas. The gain grows with the number of antennas, as discussed in Subsection \ref{basic_TD_section_sweeping}.

\vspace{-0.2cm}
\section{Conclusions and Future Works}\label{conclusions}
This work sheds some new light on the role of fading and diversity in WPT. Thanks to the energy harvester nonlinearity, fading and diversity are found beneficial to boost the RF-to-DC conversion efficiency. We developed a new form of signal design for WPT relying on phase sweeping transmit diversity. Transmit diversity relies on multiple antennas to induce fast fluctuations of the wireless channel and does not rely on any form of CSIT. The effect of fading and transmit diversity have been analyzed using two different types of energy harvester nonlinear models and the performance gain has been confirmed through circuit simulations and experimentation. The proposed scheme is well suited for massive Internet-of-Things (IoT) deployments for which CSIT acquisition is unpractical.
\par The work again stresses the importance and the benefits of modeling and exploiting the harvester nonlinearity, but also newly highlights that multiple antennas can be beneficial to WPT even in the absence of CSIT. This opens the door to a re-thinking of the role of multiple antennas and CSIT in WPT.   
\par An exciting future work consists in designing diversity strategies for the dual purpose of wireless information and power transfer. One way would be to smartly combine the transmit diversity scheme developed here and the conventional transmit diversity schemes used in wireless communications. Another interesting avenue is to identify whether other types of transmit diversity strategies could be beneficial to WPT. The prototype and implementation of transmit diversity with a larger number of antennas is also worth investigating.

\ifCLASSOPTIONcaptionsoff
  \newpage
\fi

\vspace{-0.2cm}
\section{Acknowledgments}
We thank B. Lavasani and National Instruments for providing some of the equipments needed to conduct the experiment.

\begin{IEEEbiography}{Bruno Clerckx} is a Reader in the Electrical and Electronic Engineering Department at Imperial College London. He received his PhD degrees from Universit\'{e} catholique de Louvain (Louvain-la-Neuve, Belgium) in 2005, and worked for Samsung Electronics (Suwon, Korea) from 2006 to 2011. His research area is communication theory and signal processing for wireless networks.
\end{IEEEbiography}
\begin{IEEEbiography}{Junghoon Kim} received his BSc in electronic and electrical engineering from Hongik University, Seoul, Korea, in 2008 and MSc in Telecommunications from University College London, UK, in 2015. From 2008 to 2014, He worked for Samsung Electronics, Suwon, Korea. He is currently a PhD student at Imperial College London. His research interests are RF energy harvesting and wireless information and power transfer.
\end{IEEEbiography}


\vspace{-0.2cm}
\begin{thebibliography}{1} 
\bibitem{Tse:2005} D. Tse and P. Viswanath, \emph{Fundamentals of Wireless Communication}. Cambridge, U.K.: Cambridge Univ. Press, 2005.
\bibitem{OptBehaviour} A. Boaventura, A. Collado, N. B. Carvalho, A. Georgiadis, ``Optimum behavior: wireless power transmission system design through behavioral models and efficient synthesis techniques,'' \emph{IEEE Microwave Magazine}, vol. 14, no. 2, pp. 26-35, March/Apr. 2013.
\bibitem{Costanzo:2016} A. Costanzo and D. Masotti, ``Smart solutions in smart spaces: Getting the most from far-field wireless power transfer,'' \emph{IEEE Microw. Mag.}, vol. 17, no. 5, pp. 30-45, May 2016.
\bibitem{Zeng:2017} Y. Zeng, B. Clerckx, and R. Zhang, ``Communications and Signals Design for Wireless Power Transmission,'' \emph{IEEE Trans. Commun.}, Vol 65, No 5, pp 2264 – 2290, May 2017.
\bibitem{Xu_Zhang:2014} J. Xu and R. Zhang, ``Energy beamforming with one-bit feedback,'' \emph{IEEE Trans. on Sig. Proc.}, vol. 62, no. 20, pp. 5370-5381, Oct., 2014.
\bibitem{Ku:2017} M.-L. Ku, Y. Han, B. Wang, and K. J. R. Liu, ``Joint Power Waveforming and Beamforming for Wireless Power Transfer,'' \emph{IEEE Trans. on Sig. Proc.}, vol. 65, no. 24, pp. 6409-6422, Dec. 2017.
\bibitem{Trotter:2009} M. S. Trotter, J. D. Griffin, and G. D. Durgin, ``Power-Optimized Waveforms for Improving the Range and Reliability of RFID Systems,'' in \emph{Proc. 2009 IEEE International Conference on RFID}.
\bibitem{Clerckx:2016b} B. Clerckx and E. Bayguzina, ``Waveform Design for Wireless Power Transfer,'' \emph{IEEE Trans. on Sig. Proc.}, Vol. 64, No. 23, Dec 2016.
\bibitem{Huang:2017} Y. Huang and B. Clerckx, ``Large-Scale Multiantenna Multisine Wireless Power Transfer,'' \emph{IEEE Trans. on Sig. Proc.}, vol. 65, no. 21, pp 5812-5827, Nov. 2017.
\bibitem{Huang:2018} Y. Huang and B. Clerckx, ``Waveform Design for Wireless Power Transfer with Limited Feedback,'' \emph{IEEE Trans. on Wireless Commun.}, vol. 17, no. 1, pp 415-429, Jan. 2018.
\bibitem{Clerckx:2017} B. Clerckx and E. Bayguzina, ``A Low-Complexity Adaptive Multisine Waveform Design for Wireless Power Transfer,'' \emph{IEEE Ant. and Wireless Prop. Letters}, Vol. 16, pp 2207 – 2210, 2017.
\bibitem{Moghadam:2017} M. R. V. Moghadam, Y. Zeng, and R. Zhang, ``Waveform Optimization for Radio-Frequency Wireless Power Transfer,'' \textit{IEEE SPAWC} 2017.
\bibitem{Clerckx:2018b} B. Clerckx, ``Wireless Information and Power Transfer: Nonlinearity, Waveform Design and Rate-Energy Tradeoff,'' \emph{IEEE Trans. on Sig. Proc.}, vol 66, no 4, pp 847-862, Feb. 2018.
\bibitem{Varasteh:2017} M. Varasteh, B. Rassouli, and B. Clerckx, ``Wireless Information and Power Transfer over an AWGN channel: Nonlinearity and Asymmetric Gaussian Signaling,'' \emph{IEEE Info. Theory Workshop 2017}.
\bibitem{Varasteh:2017b} M. Varasteh, B. Rassouli, and B. Clerckx, ``On Capacity-Achieving Distributions for Complex AWGN Channels Under Nonlinear Power Constraints and their Applications to SWIPT,'' arXiv:1712.01226.
\bibitem{Clerckx:2018c} B. Clerckx, R. Zhang, R. Schober, D. W. K. Ng, D. I. Kim, and H. V. Poor, “Fundamentals of Wireless Information and Power Transfer: From RF Energy Harvester Models to Signal and System Designs,” submitted to \emph{IEEE Journal on Sel. Areas in Commun.}, arXiv:1803.07123.
\bibitem{Boshkovska:2015} E. Boshkovska, D. W. K. Ng, N. Zlatanov, and R. Schober, ``Practical Non-Linear Energy Harvesting Model and Resource Allocation for SWIPT Systems,'' \emph{IEEE Comm. Letters}, vol. 19, no. 12, pp. 2082-2085, Dec 2015.
\bibitem{Xu:2017} X. Xu, A. Ozcelikkale, T. McKelvey, and M. Viberg, ``Simultaneous information and power transfer under a non-linear RF energy harvesting model,'' in \emph{Proc. IEEE Int. Conf. Communications}, pp. 179-184, May 2017.
\bibitem{Alevizos:2018} P. N. Alevizos and A. Bletsas, ``Sensitive and nonlinear far field RF energy harvesting in wireless communications,'' \emph{IEEE Trans. Wireless Commun.}, vol. 17, no. 6, pp. 3670-3685, June 2018.
\bibitem{Viswanath:2002} P. Viswanath, D.N.C. Tse, and R. Laroia, ``Opportunistic Beamforming using Dumb Antennas,'' \emph{IEEE Trans. Inform. Theory}, vol. 48, no. 6, pp. 1277–1294, Jun. 2002.
\bibitem{Hiroike:1992} A. Hiroike, F. Adachi, and N. Nakajima, ``Combined Effects of Phase Sweeping Transmitter Diversity and Channel Coding,'' \emph{IEEE Trans. Veh. Technol.}, vol. 41, no. 2, pp. 170–176, May 1992.
\bibitem{Knopp:1995} R. Knopp and P. Humblet, ``Information capacity and power control in single cell multiuser communications,'' in \emph{IEEE ICC 95}, Jun. 1995.
\bibitem{Xia:2015} M. Xia and S. Aissa, ``On the Efficiency of Far-Field Wireless Power Transfer,'' \emph{IEEE Trans. on Sig. Proc.}, vol. 63, no. 11, pp. 2835–2847, Jun. 2015.
\bibitem{SMS7630} Skyworks, ``SMS7630 Series Schottky Diode Datasheet,'' available \url{http://www.skyworksinc.com/uploads/documents/Surface_Mount_Schottky_Diodes_200041AD.pdf}
\bibitem{Alevizos:2018a} P. N. Alevizos, G. Vougioukas, and A. Bletsas, ``Nonlinear energy harvesting models in wireless information and power transfer,'' https://arxiv.org/pdf/1802.09994
\bibitem{Kim:2017}	J. Kim, B. Clerckx, and P. D. Mitcheson, ``Prototyping and Experimentation of a Closed-Loop Wireless Power Transmission with Channel Acquisition and Waveform Optimization,'' \emph{IEEE WPTC 2017}.

\end{thebibliography}
\end{document}